\PassOptionsToPackage{numbers,sort&compress}{natbib}
\documentclass[final,5p,times,twocolumn]{elsarticle}

\usepackage{amssymb}
\usepackage{lipsum}

\usepackage{amsmath}
\usepackage{mathtools}
\usepackage{multirow}
\usepackage[table,xcdraw]{xcolor}
\usepackage{colortbl}
\usepackage[normalem]{ulem}
\usepackage{threeparttable}
\usepackage{booktabs}
\usepackage{dblfloatfix}
\usepackage{graphicx}
\usepackage{caption}
\usepackage{subfig}        
\usepackage{dblfloatfix}
\usepackage{float}
\usepackage{xcolor}

\PassOptionsToPackage{hyphens}{url}
\usepackage{xurl}
\usepackage{hyperref}
\useunder{\uline}{\ul}{}

\usepackage{booktabs}
\usepackage{tabularx}
\usepackage{array}
\usepackage{makecell}
\usepackage{threeparttable}
\usepackage{cuted}
\usepackage{capt-of}

\usepackage{booktabs}          
\usepackage[version=4]{mhchem} 
\usepackage{threeparttable}    

\newcolumntype{M}[1]{>{\raggedright\arraybackslash}m{#1}}
\newcolumntype{Y}{>{\raggedright\arraybackslash}X}

\journal{Journal of Materials Processing Technology}

\begin{document}

\begin{frontmatter}

\title{Plasma Etch Process Optimization for Photonic-Grade Diamond-on-Insulator Substrates and Thickness Evaluation using Colorimetry}

\author[first,third]{Tianyin Chen} 
\author[first,second,third]{Alessio Miranda} 
\author[first,second]{Leyla Rami} 

\author[first,second]{Ryoichi Ishihara} 
\author[first]{Salahuddin Nur}

\affiliation[first]{organization={Department of Quantum Computer Engineering, Faculty of Electrical Engineering
Mathematics and Computer Science,},
            addressline={Delft University of Technology}, 
            city={Delft},
            country={The Netherlands}}

            \affiliation[second]{organization={QuTech},
            addressline={Delft University of Technology}, 
            city={Delft},
            country={The Netherlands}}
\affiliation[third]{These authors contributed equally to this work}

\begin{abstract}
Diamond color-center qubits integrated with photonic circuits can be initialized, manipulated, entangled, and read individually with high fidelity, which makes this framework ideal for implementing large-scale, modular quantum computers, quantum networks, and distributed quantum sensing systems. However, the limited size of heteroepitaxially grown single-crystal diamond and photonic-grade diamond-on-insulator (DOI) substrates poses challenges for their integration with existing manufacturing processes. In this work, we have developed a plasma etch recipe for efficiently etching direct-bonded (100) single-crystal diamond membranes ($<$ 50 $\mu$m) to produce large-area, thin-film DOI substrates, and we demonstrate the fabrication of free-standing photonic chiplets using the DOI substrate produced with our recipe. The plasma etch recipe, optimized on an Inductively Coupled Plasma Reactive Ion Etching (ICP RIE) system, preserves diamond bonding, provides sufficient micromasking and surface-quality control, and enables the manufacture of thin-film DOI substrates. Using this recipe, we have thinned down a 10 $\mu$m diamond plate bonded to a SiO$_2$/Si substrate and produced a photonic grade (thickness $\leq$ 300 nm) DOI substrate. We have achieved a DOI film of uniform thickness around 300 nm over an area of 0.5 × 0.5 mm$^2$ and surface roughness below 0.5 nm, while preserving the bonding interface intact. We fabricate diamond photonic chiplets with this DOI substrate using a standard two-step lithography process, without needing any complex thin film transfer, diamond under-etching steps, or forming any pedestal geometries.
We also present a colorimetric study of the visibility of a diamond layer on SiO$_2$ and quantify color differences across various thicknesses in common colorimetric spaces. Finally, this analysis has been utilized for an automatic extrapolation of the diamond thickness from standard optical microscope images with a resolution of 5nm, which agrees well with White Light Interferometer (WLI) thickness measurement results. This large-area, thin diamond-on-insulator (DOI) substrate, combined with the colorimetric thickness evaluation technique, provides an effective fabrication platform and reliable thickness validation method for scalable manufacturing of diamond nanophotonic devices, thereby opening a path toward realizing large-scale, integrated quantum systems.
\end{abstract}




\end{frontmatter}




\section{Introduction}
\label{introduction}

Diamond has become one of the leading material platforms for quantum technologies~\cite{knaut2024entanglement-459, xu2022review-b98, Aharonovich2011cu, 10.5555/3312180, Atature2018hh, MaxRuf_Network2021} and extreme-environment devices. This broad relevance arises from its combination of an ultra-wide bandgap ($\approx 5.47$~eV), exceptional thermal conductivity ($\sim$2200~W$\cdot$m$^{-1}$$\cdot$K$^{-1}$, among the highest of natural materials), high mechanical stiffness, optical transparency from the UV to the mid-IR, and chemical robustness~\cite{xu2022review-b98, PEREZ2020108154, donato2020diamond-d7a, nano14050460, liang2024direct-4cd, jne2040032, khanna_harsh2023, 9798392}. In conventional device domains, these properties underpin low-loss photonics, including Raman lasers and nonlinear optics, optoelectronic components, and extreme-environment electronics with high breakdown fields and efficient heat dissipation~\cite{latawiec2015onchip-634, hausmann2014diamond-521, PEREZ2020108154, donato2020diamond-d7a, nano14050460, liang2024direct-4cd, jne2040032, khanna_harsh2023, 9798392}. Beyond its bulk material properties, diamond hosts atom-like point defects, or color centers, including the nitrogen-vacancy (NV) center and group-IV split-vacancy centers such as SiV, GeV, and SnV. These defects provide optically addressable electronic spins with long coherence and spin-selective optical transitions~\cite{Atature2018hh, katsumi_recent_2025, bradac2019quantum-710, MaxRuf_Network2021, janitz_cavity_2020, pezzagna2021quantum-9b0}. As a result, diamond color centers have become attractive building blocks for quantum computing, quantum communication networks, and distributed quantum sensing systems. They have already enabled diamond-based single-photon sources \cite{katsumi_recent_2025, mizuochi2012electrically-6ee, Aharonovich2011cu, torun2026super-acb} quantum sensing systems \cite{Schirhagl2014io, levine2019principles-5ef, wang2025fully-515, varveris2026toward-789,  katsumi_recent_2025}, long-lived quantum memories \cite{fuchs2011quantum-782, Maurer2012kg, bradley2022robust-c33}, and landmark demonstrations including spin–photon entanglement and multi-node quantum networking \cite{ bradley2022robust-c33, knaut2024entanglement-459, pompili2021realization-49b, MaxRuf_Network2021, stas2026entanglementassisted-32a}.

Realizing practical systems based on these solid-state qubits requires efficient spin--photon interfaces and low-loss on-chip routing~\cite{bradac2019quantum-710, janitz_cavity_2020, katsumi_recent_2025, shandilya2022diamond-be6, Awschalom2018ic, ishihara3DIntegrationTechnology2021, wan2020largescale-f02}. Embedding color centers in nanophotonic structures, such as waveguides, microdisks/rings, and photonic-crystal (PhC) cavities, enhances emission into useful optical modes, increases collection efficiency, and enables cavity-QED regimes in which Purcell enhancement accelerates coherent photon generation and can improve entanglement rates~\cite{bradac2019quantum-710, janitz_cavity_2020, ding2025purcellenhanced-76b, fischer2025spinphoton-f13, katsumi_recent_2025, Awschalom2018ic, MaxRuf_Network2021, ding_high-q_2024}. Recent progress highlights this opportunity: thin-film diamond PhC cavities with record visible-wavelength quality factors of $Q\approx 1.6$--$1.8 \times 10^5$ have been demonstrated, and group-IV centers such as SnV$^-$ have shown strong cavity-enhanced emission with large $\beta$-factors in nanostructures~\cite{rugar_quantum_2021, janitz_cavity_2020, fischer2025spinphoton-f13, katsumi_recent_2025, MaxRuf_Network2021, lee2026quantum-3b8, ding2025purcellenhanced-76b, ding_high-q_2024, riedel2026scalable-5f4}.

However, the transition from one-off diamond devices to scalable integrated circuits remains constrained by materials, substrate availability, and processing realities. High-quality, uniform thin single-crystal diamond (SCD) layers are still difficult to realize at scale. Commercial high-purity SCD is typically available only as small plates (mm–cm scale), while wafer-scale SCD by direct heteroepitaxy remains challenging due to nucleation, lattice and thermal-expansion mismatch, residual strain, mosaicity, and high dislocation densities~\cite{ rani2020recent-b6c, hill2018thin-06f, piracha_scalable_2016, shandilya2022diamond-be6, uwihoreye2024recent-de5, fischer2025spinphoton-f13, katsumi_recent_2025, lee2026quantum-3b8, ding_high-q_2024, riedel2026scalable-5f4, bensalah2016mosaicity-11c, mandal2021nucleation-68c, chen_hydrophilic_2025}. Consequently, much of high-performance SCD nanophotonics has relied either on directly machining free-standing structures from bulk diamond, using angled/quasi-isotropic etching or focused-ion-beam methods, or on fabricating and transferring thin membranes derived from bulk through ion-slicing, overgrowth, or smart-cut-like approaches. These routes have enabled state-of-the-art proof-of-principle devices, including high-Q cavities, waveguides, and resonators. Nevertheless, they remain limited by throughput, design flexibility, surface damage, membrane area, variability in surface quality and thickness uniformity, and roughness on the back surface and pedestal formation. These parameters directly affect optical scattering loss, cavity yield, resonance reproducibility, spectral matching to color-center transitions, and ultimately the scalability of diamond quantum photonic circuits~\cite{Burek2014bj, rani2020recent-b6c, lee2013fabrication-738, guo_tunable_2021, guo2024directbonded-8fc, ding_high-q_2024, hill2018thin-06f, shandilya2022diamond-be6, katsumi_recent_2025}.

In conventional bulk-diamond free-standing fabrication routes, such as quasi-isotropic oxygen etching, the undercut is governed by crystal-plane-dependent etch profiles and by the local geometry of the patterned structure~\cite{Khanaliloo_2015, rani2020recent-b6c, mi_integrated_2020}. As a result, quasi-isotropic etching can naturally form pedestal-supported suspended devices, as reported for monolithic diamond microdisks, where an hourglass-shaped pedestal remains beneath the released disk~\cite{Khanaliloo_2015}. For more complex layouts, such as cross-waveguide or intersecting-waveguide structures \cite{miranda2026cross-c44}, this geometry-dependent undercut can make complete release more difficult, as different arms and junction regions may provide different etch-access paths and can leave residual support regions unless the release is carefully engineered. In contrast, the DOI process uses the buried SiO$_2$ as a sacrificial layer, enabling a more direct release, when necessary, of frame-supported diamond photonic chiplets for pick-and-place type heterogeneous integration.


In this work, we address the manufacturing bottlenecks by developing and validating a plasma-etch process for photonic-grade DOI starting from direct-bonded (100) SCD membranes ($<50~\mu$m). We optimize an ICP-RIE recipe that alternates etching and surface-cleaning steps to (i) suppress debonding during aggressive thinning, (ii) minimize micromasking and nanopillar formation, and (iii) deliver smooth, uniform films suitable for nanophotonics. Using this process, we thin a 10~$\mu$m bonded diamond plate on SiO$_2$/Si down to $\approx$300~nm while maintaining sub-nanometre roughness over a $0.5 \times 0.5$~mm$^2$ area and preserving the bonding interface, meeting practical targets for integrated diamond waveguides and cavities.

Furthermore, to support rapid and accessible metrology during thinning, we develop a colorimetry-based thickness evaluation method tailored to diamond/SiO$_2$/Si stacks. By analyzing color distances across standard color spaces for different diamond/SiO$_2$ thickness combinations, we establish a robust mapping from optical micrographs to diamond film thickness with $\approx$5~nm resolution. This mapping is validated against white-light interferometry (WLI). The method provides a low-cost, microscope-based gauge for tracking the thinning process, shortening plasma-etch optimization cycles and enabling large-area uniformity checks without specialized thickness-metrology instrumentation.

Taken together, the optimized plasma thinning process, the frame-supported photonic chiplet demonstration and the colorimetric thickness evaluation provide a practical route to photonic-grade DOI substrates with the surface quality and thickness control required for reproducible diamond photonic device fabrication. Although the primary motivation is to advance diamond quantum nanophotonics towards scalable, heterogeneously integrated, and ultimately 3D-integrated systems, the same DOI platform and metrology are also relevant to classical diamond photonics, optoelectronics, and high-power or extreme-environment diamond electronics. The processing methodology, therefore, extends beyond quantum applications and provides a broadly applicable route for manufacturing thin, device-ready single-crystal DOI substrate.


\section{Current status of thin-film diamonds}
\label{sec:thin_film}


\begin{table*}[t]
\vspace{-0.5\baselineskip}
\centering
\scriptsize
\setlength{\tabcolsep}{2.8pt}
\renewcommand{\arraystretch}{1.12}

\begin{threeparttable}
\captionof{table}{Brief comparison of representative thin-film diamond and diamond-on-insulator (DOI) approaches.}
\label{tab:doi_comparison}

\begin{tabularx}{\textwidth}{
M{0.105\textwidth}
Y
Y
Y
Y
Y
}
\toprule
\textbf{Category} &
\textbf{\makecell[l]{High group,\\ Chicago \cite{guo2024directbonded-8fc}}} &
\textbf{\makecell[l]{High/\\ Lon\v{c}ar \cite{ding_high-q_2024}}} &
\textbf{\makecell[l]{Maletinsky group* \cite{corazza2025homogeneous-a42}}} &
\textbf{\makecell[l]{IonQ\\ platform* \cite{riedel2026scalable-5f4}}} &
\textbf{Our work} \\
\midrule

\textbf{Technique} &
Direct-bonded released membranes &
Bonded thin-film diamond cavity platform &
Thinning-down etch; free-standing membranes &
\vspace{0.5em} Thermocompression-bonded membrane transfer + thinning &
Thinning-down etch on directly bonded DOI \\

\textbf{\makecell[l]{Area /\\ roughness}} &
$200 \times 200~\mu\mathrm{m}^2$ / NR &
$\sim 200 \times 200~\mu\mathrm{m}^2$ / $< 0.3$ nm &
mm-scale / $< 0.2$ nm &
\vspace{0.5em} 0.5 mm hexagonal membranes / $< 0.3$ nm &
$500 \times 500~\mu\mathrm{m}^2$ / $< 0.5$ nm \\

\textbf{\makecell[l]{Thickness /\\ variation}} &
Down to 10 nm / nm-scale variation &
160 nm / $\sim 1$ nm variation &
Down to 70 nm / $< 0.35$ nm/$\mu$m &
\vspace{0.5em} 160 nm / post-bond thin films demonstrated &
$\sim 300$ nm / $< 0.08$ nm/$\mu$m \\

\textbf{\makecell[l]{Bonding /\\ support}} &
Directly bonded to Si, $\mathrm{SiO_2}$, fused silica, sapphire, $\mathrm{LiNbO_3}$ &
Bonded to $\mathrm{SiO_2}$/Si; Cr/Au frame used &
Free-standing membrane &
\vspace{0.5em} Gold thermocompression bonding to semiconductor substrate &
Plasma-activated hydrophilic bonding to $\mathrm{SiO_2}$/Si \\

\textbf{Key point} &
Multi-substrate heterogeneous bonding &
High-$Q$ visible cavities &
Best recent free-standing uniformity &
\vspace{0.5em} Scalable membrane transfer and integration &
Simple bonded-on-Si DOI with in-line colorimetry \\

\textbf{\makecell[l]{Relative position\\ vs this work}} &
True bonded stack, but much smaller area &
Strong device result, but smaller area and added metal-frame complexity &
Excellent area and uniformity, but not a true DOI stack &
Scalable, but uses metal-interlayer thermocompression and a more complex transfer route &
\vspace{0.5em} True DOI on $\mathrm{SiO_2}$/Si; larger photonic-grade bonded area; no metal interlayer; preserved buried interface; simple, scalable process \vspace{0.5em}\\

\bottomrule
\end{tabularx}

\begin{tablenotes}[flushleft]
\footnotesize
\item DOI - diamond-on-insulator, NR - not reported.
\item * Not a true DOI stack.
\end{tablenotes}
\end{threeparttable}
\vspace{-0.4\baselineskip}
\end{table*}



To realize large-scale integrated diamond photonic circuits, a suitable substrate architecture is required. A leading monolithic candidate is the diamond-on-insulator (DOI) platform, consisting of a thin, photonic-grade SCD membrane bonded to a low-refractive-index substrate, typically silicon dioxide (SiO$_2$) on a silicon handle wafer~\cite{chen_hydrophilic_2025, guo2024directbonded-8fc, ding_high-q_2024, katsumi_recent_2025}. This architecture is analogous to the silicon-on-insulator (SOI) platform that transformed microelectronics and silicon photonics~\cite{bruel1995silicon-6e7, bogaerts2004basic-165, plobl2000silicononinsulator-fb7}. In DOI, the strong vertical refractive-index contrast between diamond ($n\approx 2.4$) and SiO$_2$ ($n_{SiO_2}\approx 1.45$) enables tight optical confinement in the diamond layer, supporting compact waveguides and high-Q cavities. This represents a shift from proof-of-principle experiments on individual color centers or released devices toward engineered, scalable, and integrated diamond quantum photonic systems. Analogous to SOI in silicon photonics~\cite{bogaerts2004basic-165, vlasov2004losses-6a2, ashida2018photonic-939, Nur2019ka, reed2010silicon-799}, DOI can avoid complex diamond undercut steps~\cite{ Burek2014bj, khanaliloo2015singlecrystal-e3b, Khanaliloo_2015, sipahigil_integrated_2016, mi_integrated_2020, rani2020recent-b6c, pasini2024nonlinear-cc5, katsumi_recent_2025} and enable planar processing of high-index-contrast waveguides, 1D/2D PhC cavities, couplers, and more complex photonic circuits. It also provides a natural route to dense integration with detectors, electrodes, and heterogeneous materials. Recent thin-film DOI-like platforms have already demonstrated high-Q cavities and uniform device arrays~\cite{ guo2024directbonded-8fc, ding_high-q_2024, katsumi_recent_2025}, provided that the diamond film offers: (i) a few-hundred-nanometre thickness suitable for single-mode guiding, (ii) low surface roughness, typically below $\sim$1~nm over device areas, to suppress optical scattering, and (iii) low total-thickness variation (TTV) to maintain spectral uniformity across a chip. 

Epitaxial growth of SCD on non-diamond substrates remains highly challenging. Although heteroepitaxial diamond has been demonstrated on specialized crystalline templates, most notably Ir-based buffer stacks, these approaches require complex lattice and thermal-expansion management and still suffer from mosaicity, residual strain, and dislocations. In contrast, direct growth or deposition of a photonic-grade SCD film on low-index oxide substrates such as SiO$_2$ has not been demonstrated, to the best of the present literature. Reported diamond growth on SiO$_2$ or fused silica typically yields microcrystalline, nanocrystalline, or polycrystalline diamond rather than photonic-grade SCD~\cite{jiang2002growth-c8d, rani2020recent-b6c, mi_integrated_2020, katsumi_recent_2025, bensalah2016mosaicity-11c, mandal2021nucleation-68c, wang2022heteroepitaxy-a86, chen_hydrophilic_2025}. Therefore, bonding of pre-existing SCD plates remains so far the most practical route to attach diamond to low-index oxide materials and realize DOI substrates~\cite{rani2020recent-b6c, mi_integrated_2020, katsumi_recent_2025, hill2018thin-06f, guo2024directbonded-8fc, ding_high-q_2024, chen_hydrophilic_2025}. Among the available bonding approaches, hydrophilic direct bonding is particularly attractive because it enables direct bonding of (100) SCD plates to plasma-enhanced chemical vapor deposition (PECVD)-grown SiO$_2$/Si substrates at low temperature and atmospheric conditions, as demonstrated in our previous work~\cite{chen_hydrophilic_2025}. However, producing large-area, photonic-grade DOI substrates with film thicknesses near or below 300~nm remains challenging to date ~\cite{rani2020recent-b6c, mi_integrated_2020, katsumi_recent_2025, hill2018thin-06f, guo2024directbonded-8fc, ding_high-q_2024, chen_hydrophilic_2025}. 

Several groups have explored photonic-grade SCD membrane or film fabrication using processes similar to the smart-cut technology~\cite{lee2013fabrication-738, piracha_scalable_2016, rani2020recent-b6c, guo_tunable_2021, guo2024directbonded-8fc, chretien_transfer_2025, katsumi_recent_2025}. These approaches can produce high-quality membranes, but they typically require many demanding and expensive steps, including ion implantation to create a buried damage layer, SCD overgrowth, membrane lift-off, transfer printing, and complex membrane bonding. As a result, the produced membranes have so far remained limited in area ($\leq$ 250 $\mu$m)  and scalability for large photonic circuits and 3D-integrated on-chip quantum systems~\cite{guo2024directbonded-8fc, guo_tunable_2021, ding_high-q_2024, katsumi_recent_2025}. Only recently, a membrane area of ~ 500 x 500 $\mu$m$^2$ has been reported using process steps such as ion implantation, membrane liftoff, high-quality overgrowth, and flip-chip bonding with metal pads \cite{riedel2026scalable-5f4}.  However, this process produces free-standing membranes for photonic structures rather than a conventional DOI substrate, which limits its compatibility and scalability with all-diamond photonics and subsequent 3D-integration steps~\cite{riedel2026scalable-5f4}.

The central material-processing challenge is, therefore, to thin and planarize direct-bonded or transferred SCD membranes into photonic-grade DOI films while preserving both surface and interface quality. Dry etching is indispensable for this step~\cite{mi_integrated_2020}, but oxygen-based inductively coupled plasma reactive ion etching (ICP-RIE) of diamond is sensitive to micromasking from particulates or redeposition, can generate nanopillars and pits, and may induce roughness that degrades optical performance \cite{mi_integrated_2020, ruf_optically_2019, hicks_diamond_2019, toros_reactive_2020, heupel_fabrication_2020, Burek2012}. Process windows that combine efficient material removal with smoothing or cleaning steps, such as cyclic O$_2$ and Ar/Cl$_2$ plasmas, can mitigate these effects and achieve sub-nanometer roughness suitable for high-finesse optics and cavity-QED~\cite{mi_integrated_2020, ruf_optically_2019, hicks_diamond_2019, hicks_optimizing_2019, toros_reactive_2020, heupel_fabrication_2020, Burek2012}. In parallel with surface quality, bonding interfaces are critical for DOI reliability, thermal budgets, and 3D integration. Hydrophilic direct bonding of diamond to SiO$_2$/Si is attractive because it avoids metallic or polymeric interlayers that can introduce absorption, stress, or contamination detrimental to qubit coherence and photonic loss~\cite{katsumi_recent_2025,  guo2024directbonded-8fc, chen_hydrophilic_2025}. Recent work has demonstrated direct bonding of (100) diamond to deposited SiO$_2$ at low temperature, with interfacial chemistry and pre-bond surface preparation, including plasma/piranha activation and roughness control, directly correlated with bonding yield and shear strength~\cite{chen_hydrophilic_2025}.

\begin{figure*}[b]
    \centering
    \includegraphics[width=0.8\textwidth]{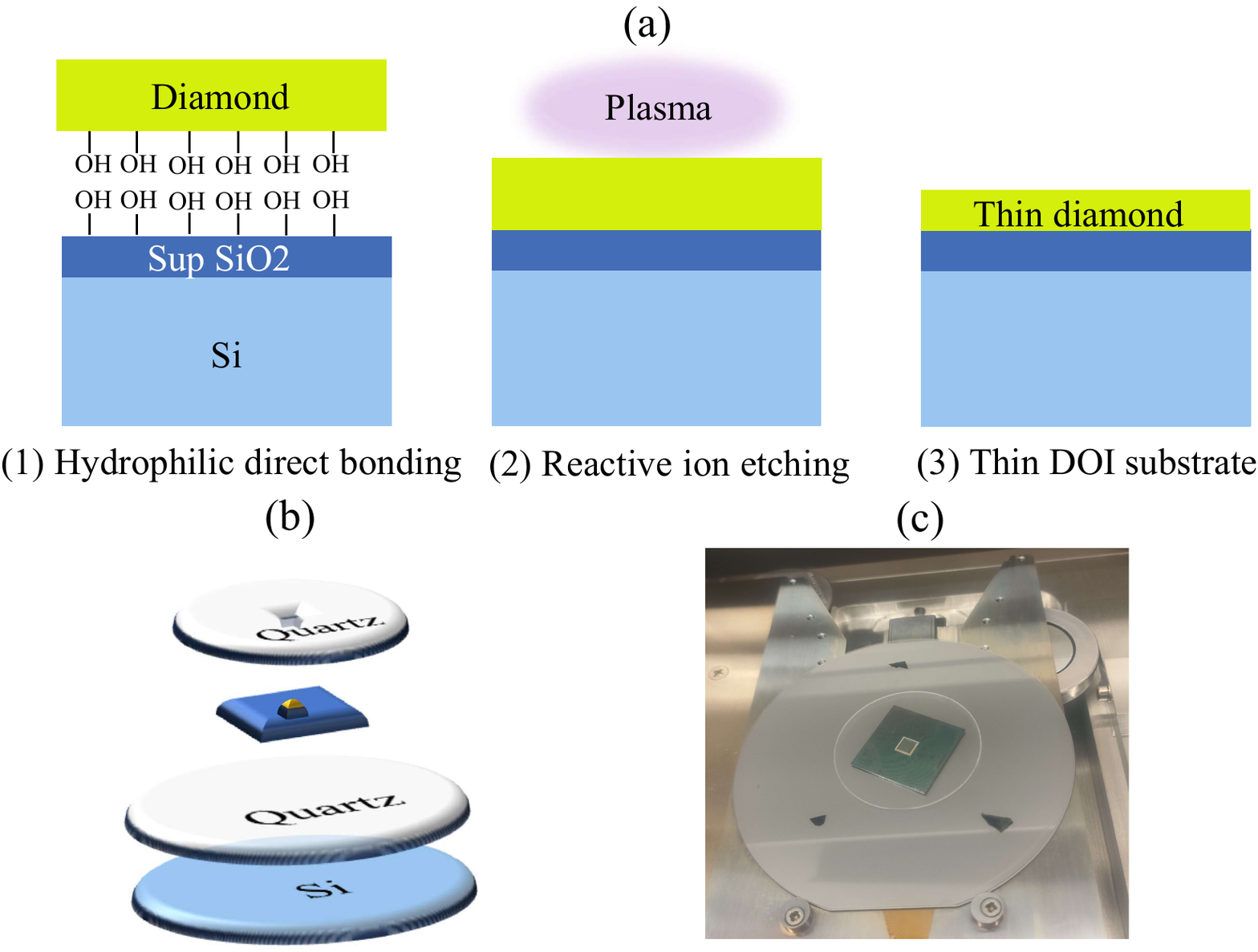}
    \caption{(a) Schematic of the diamond thinning process. Starting with a commercial diamond plate and Si wafer, we directly bond the two plates to form the initial DOI substrate. The diamond plate thickness is then reduced by reactive ion etching, yielding a thin and smooth DOI substrate. (b) Optimized ICP-RIE carrier-wafer and mask concept. The diamond-on-insulator sample is the specimen to be etched, and a quartz mask is applied on top of the sample to minimize micromasking. The DOI substrate is mounted on a double-layer carrier-wafer stack during etching. (c) DOI sample loaded in the RIE tool (Plasmalab  System 100, Oxford Instruments) for etching.}
    \label{figure_schematic}
\end{figure*}

Table~\ref{tab:doi_comparison} compares the present work with recent thin-film diamond approaches. Unlike free-standing membranes or multi-step transferred membrane schemes, our process specifically targets a true direct-bonded DOI platform on SiO$_2$/Si. The key distinction is a carrier-compatible route that combines sub-mm-scale photonic-grade DOI area, deep thinning to $\sim$300~nm, preservation of the buried diamond--oxide interface, and the absence of metal or intermediate bonding layers. The resulting platform is therefore closer to a practical DOI substrate than to a released membrane demonstrator. To validate the usability of the manufactured DOI layer, we further fabricate frame-supported diamond photonic chiplets, in which waveguides are retained within a surrounding mechanical frame. This geometry provides a practical building block for heterogeneous integration of large-scale diamond quantum photonic systems.

Looking ahead, 3D integration that vertically stacks diamond qubit layers, photonics, and cryo-CMOS control electronics is regarded as a promising route to modular quantum processors and network nodes~\cite{ishihara3DIntegrationTechnology2021}. A DOI foundation layer is well suited to such stacks because it provides a planar photonic base that can co-integrate with superconducting, nonlinear, or piezoelectric films for frequency conversion and phonon-mediated interfaces, as recently explored in hybrid thin-film diamond platforms.

\section{Photonic-grade DOI fabrication process}
\label{sec:process_description}

ICP-RIE is one of the most widely used approaches for thinning and structuring single-crystal diamond because it decouples plasma density and ion bombardment energy, allowing the etch rate, anisotropy, and surface modification to be tuned independently \cite{HICKS2019,toros_reactive_2020}.
Early studies based on oxygen-rich plasmas established that diamond can be etched at high
rates under ICP conditions, while later work showed that long oxygen-based etches often
promote micromasking, nanopillar formation, and surface roughening if particles or
redeposited material are present on the surface
\cite{hwang_new_2004,enlund_anisotropic_2005,hausmann_fabrication_2010,toros_reactive_2020}.
Chlorine-based plasmas, particularly Ar/Cl$_2$, were subsequently introduced as a
complementary route because they can smooth diamond surfaces and reduce the
preferential roughening frequently observed during long O$_2$ etches
\cite{lee_etching_2008,HICKS2019, ruf_optically_2019, heupel_fabrication_2020}. Therefore, more recent work has converged 
on cyclic processes that alternate an oxygen-rich etching step with a chlorine-based
cleaning/smoothing step to combine acceptable removal rates with reduced micromasking
and improved surface quality \cite{HICKS2019,heupel_fabrication_2020}.

\begin{table*}[hb]
\centering
\scriptsize
\setlength{\tabcolsep}{3.8pt}
\renewcommand{\arraystretch}{1.12}
\begin{threeparttable}
\captionof{table}{Representative literature ICP-RIE routes relevant to DOI thinning and surface-quality control.}
\label{tab:etch_lit_overview}
\begin{tabularx}{\textwidth}{
M{0.17\textwidth}
M{0.15\textwidth}
M{0.18\textwidth}
Y
M{0.12\textwidth}
Y
}
\toprule
\textbf{Study} &
\textbf{Chemistry} &
\textbf{Typical operating window} &
\textbf{Main purpose} &
\textbf{Reported outcome} &
\textbf{Relevance here} \\
\midrule
Hwang et al.\ \cite{hwang_new_2004} &
Ar/O$_2$ &
ICP 700--1000 W; bias 200 W; Ar/O$_2$ = 30/90 sccm &
High-rate anisotropic ICP etching &
Etch rate up to 40~$\mu$m/h &
\vspace{0.5em} Illustrates why high-power oxygen-based recipes are attractive starting points for fast diamond removal \vspace{0.5em}\\

Enlund et al.\ \cite{enlund_anisotropic_2005} &
Ar/O$_2$ &
2.5 mTorr; RF 600 W &
Anisotropic etching of single-crystal CVD diamond &
Etch rate $>$200 nm/min; smooth vertical features &
\vspace{0.5em} Shows the effectiveness of oxygen/argon ICP etching, but not its compatibility with bonded DOI \vspace{0.5em}\\

Lee et al.\ \cite{lee_etching_2008} &
Ar/Cl$_2$ & \vspace{0.5em}  There is no optimized recipe. Recipe window:
5 mTorr; Ar/Cl$_2$ = 25/40 sccm; ICP coil power 100--900 W, constant ICP platen power of 300 W &
Surface smoothing / isotropic chlorine-based etch &
rms roughness reduced from 0.53 to 0.19 nm after 10 min &
\vspace{0.5em} Motivates the use of Ar/Cl$_2$ as a cleaning/smoothing step rather than the main high-rate etch \vspace{0.5em}\\

Hicks et al.\ \cite{hicks_diamond_2019} &
Cyclic Ar/O$_2$--Ar/Cl$_2$ & \vspace{0.5em}\vspace{0.5em}  Ar/O$_2$: 20\% O$_2$ in Ar (50 sccm Ar), 5 mTorr, ICP power 100–200 W, platen power 300 W
Ar/Cl$_2$: 5 mTorr, ICP power 100 W, platen power 250 W, 15 sccm Ar, 7 sccm Cl$_2$,
Cycle: 2 min Ar/O$_2$ + 20 s Ar/Cl$_2$ cycles &
Deep etching with reduced micromasking &
10.6~$\mu$m depth; 45.1 nm/min; 0.47 nm rms; near-zero micromasking \vspace{0.5em}&
Provides the key literature precedent for cyclic etching during long diamond-thinning runs \vspace{0.5em}\\

Ruf / Heupel et al.\ \cite{ruf_optically_2019,heupel_fabrication_2020} &
Alternating Ar/Cl$_2$ and O$_2$; optional soft Ar/Cl$_2$ &
Ar/Cl$_2$: RF/ICP = 200/500 W; O$_2$: 90/1100 W; soft Ar/Cl$_2$: 40/200 W &
Membrane thinning, strain-relief etch, and roughness reduction &
\vspace{0.5em} Quartz-supported membrane processing; low-bias Ar/Cl$_2$ used for additional smoothing \vspace{0.5em}&
Provides the standard recipe family used as the initial reference point in this work \\
\bottomrule
\end{tabularx}
\begin{tablenotes}[flushleft]
\footnotesize
\item The literature table is intended to motivate the recipe-development logic rather than to provide an exhaustive recipe catalog.
\end{tablenotes}
\end{threeparttable}
\end{table*}
\vspace{0.2em}

For directly bonded DOI substrates \cite{chen_hydrophilic_2025}, these general considerations
become more restrictive. In addition to maintaining a reasonable etch rate and a smooth
diamond surface, the process must preserve the buried bonding interface and avoid
exposing or damaging the underlying oxide/support stack. This is particularly important
when a directly bonded diamond plate is thinned from the tens-of-micrometers range to
the few-hundred-nanometer regime required for photonic-grade DOI. In the present work,
the process flow therefore combines carrier-wafer engineering, gas-chemistry optimization,
top-side masking, and cyclic etching. The overall route starts from a directly bonded DOI
stack, then progressively modifies the carrier/support design and plasma conditions to
suppress debonding and micromasking, and finally applies cyclic thinning to obtain a
$\sim$300~nm diamond film over a $0.5 \times 0.5$~mm$^2$ area.

As illustrated in Fig.~\ref{figure_schematic}, a commercially available single-crystal diamond
plate is first directly bonded onto an SiO$_2$/Si carrier \cite{chen_hydrophilic_2025}, forming the initial DOI structure.
The bonded sample is subsequently mounted for ICP-RIE thinning using a quartz-based
carrier/mask assembly designed to minimize redeposition onto the diamond surface. In the
final process flow, top-side quartz masking is used to limit plasma exposure of the
surrounding carrier wafer, thereby reducing the likelihood of micromasking originating from
carrier erosion or redeposition. The thickness evolution of the thin diamond layer is assessed by
Dektak stylus profilometry, WLI, and Raman spectroscopy,
while surface quality is tracked by optical microscopy, scanning electron microscope (SEM), and atomic force microscopy (AFM). This separation
between thickness metrology and surface-quality inspection proved useful throughout
recipe development.

The trends in the literature summarized in Table~\ref{tab:etch_lit_overview} directly motivate the
recipe strategy used in this work. Oxygen-rich ICP-RIE provides the required material removal rate, but long oxygen-based etches are susceptible to roughening and micro-masking-related defect formation. Ar/Cl$_2$, in contrast, is less suitable as the sole high-rate etch but is valuable for cleaning and smoothing \cite{ruf_optically_2019, heupel_fabrication_2020}. Therefore, process development is focused on translating this established cyclic logic from free-standing membrane fabrication to directly bonded DOI substrates, where preservation of the bonding interface
becomes an equally important constraint.



\section{Etch recipe development for thinning bonded DOI substrates}
\label{thinning_down_recipe}

In this section, we describe the development of a dry etching recipe for thinning down SCD bonded on SiO$_2$/Si substrates. As illustrated in Fig.~\ref{figure_schematic} (a), we started with a commercially available SCD plate (50-500~$\mu$m thick) and a silicon wafer. A 300~nm layer of SiO$_2$ was first deposited on the Si wafer, acting as the buried oxide layer of the DOI platform. After surface treatment of both the diamond and SiO$_2$/Si wafer, the two plates were directly bonded \cite{chen_hydrophilic_2025}. The bonded plate stack, which already formed a DOI substrate, was subsequently thinned by reactive ion etching to reduce the diamond thickness. Our goal is to fabricate a DOI substrate with a controllable thickness in the sub-micrometer range and a surface roughness below 0.5~nm, without compromising either the etch rate or the surface quality of the diamond.

A critical part of our process is dry etching, which involves preparing the carrier wafer and optimizing the etching chemistry. The carrier-wafer concept used for loading into the etch chamber is shown in Fig.~\ref{figure_schematic}(b). We selected quartz as both mask and carrier-wafer material wherever possible because it showed the least material redeposition during oxygen-plasma-based etching compared to other commonly used materials \cite{ruf_optically_2019}. This became particularly important for long etch runs, where redeposition and particle accumulation can lead to severe micromasking.


\begin{table*}[!b]
\centering
\footnotesize
\renewcommand{\arraystretch}{1.15}
\caption{DOI substrates used for etch-recipe optimization and corresponding plasma-etch parameters.}
\label{etch_recipes}
\begin{threeparttable}
\begin{tabular}{@{}lcccccc@{}}
\toprule
\textbf{Sample code and process step} & \textbf{ICP / RF (W)} & \textbf{Gas flow (sccm)} & \textbf{Pressure ($\mu$bar)} & \textbf{Carrier $T$ ($^\circ$C)} & \textbf{Duration} & \textbf{Result} \\
\midrule
Initial-DOI etch step 1 & 500 / 200 & 10 Ar / 20 \ce{Cl2} & 10 & 30 & 30 mins& Debonded \\

Initial-DOI etch step 2 & 1100 / 90 & 50 \ce{O2} & 10 & 20 & 30 mins & Debonded\\

Thick-DOI gentle etch step 1 & 180 / 30 & 30 \ce{O2} & 13 & 20 & 15 mins & Remain bonded\\
Thick-DOI gentle etch step 2 & 500 / 200 & 50 Ar / 50 \ce{O2} & 50 & 20 & 15 mins and 60 mins & Remain bonded\\

DOI-quartz etch step 1& 1000 / 90 & 50 \ce{O2} & 10 & 20 & 15 mins & Remain bonded\\
DOI-quartz etch step 2 & 500 / 200 & 10 Ar / 20 Cl\textsubscript{2} & 10 & 30 & 20 mins & Remain bonded\\
\bottomrule
\end{tabular}
\end{threeparttable}
\end{table*}


\subsection{Initial etching recipe and challenges}

We first tested an etching sequence adapted from the process reported by Ruf \textit{et al.} for producing optically coherent NV centers in micrometer-thin etched diamond membranes~\cite{ruf_optically_2019}. The corresponding parameters are listed in Table~\ref{etch_recipes} as the ``Initial-DOI etch''. In our case, the starting substrate was a $4~\mathrm{mm} \times 4~\mathrm{mm}$ single-crystal diamond plate directly bonded to a 300~nm PECVD SiO$_2$ layer on a Si handle wafer. In the reference process, the Ar/Cl$_2$ plasma is used primarily as a surface-cleaning step to remove particles and suppress micromasking, whereas the O$_2$ plasma provides the main diamond-removal step. However, when these relatively high-power plasma conditions were applied to the directly bonded DOI samples, debonding occurred during both the Ar/Cl$_2$ and O$_2$ steps. The debonding was identified optically from the appearance of interference fringes after etching, consistent with the formation of an air gap at the diamond--oxide bonding interface.


\begin{figure*}[ht!]
    \centering
    \includegraphics[width=1\textwidth]{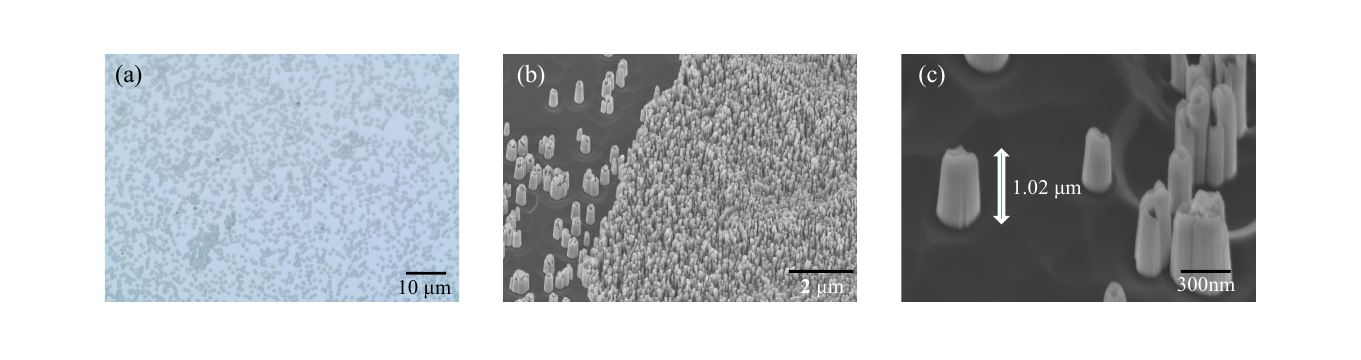}
    \caption{Observed surface damage after long etching with initial DOI etching recipe, without quartz mask. (a) optical microscope image of the surface. (b) Top-view SEM image of the diamond surface after extended etching, showing nanopillar-like structures associated with micromasking. (c) A nanopillar with a height of 1.04~$\mu$m, imaged at 60k magnification. The sample carrier was tilted by 15$^\circ$.}
    \label{figure_sem_masking}
\end{figure*}



Two possible mechanisms were considered for the observed debonding. The first was a lateral attack or under-etching of the buried SiO$_2$ layer. Because the Ar/Cl$_2$ plasma can have an isotropic component and parts of the surrounding carrier surface may be exposed during etching, the buried oxide could, in principle, be attacked from the diamond edges. However, for a $4~\mathrm{mm} \times 4~\mathrm{mm}$ bonded diamond plate, complete removal of the oxide underneath the diamond by lateral etching alone is unlikely over the time scale of the process. In addition, debonding was also observed during the O$_2$ etching step, where chemical removal of SiO$_2$ is not expected to be the dominant process. We therefore considered oxide under-etching to be insufficient by itself to explain the observed failure.

The second and more plausible mechanism would be plasma-induced thermomechanical stress. Under high-power plasma conditions, the Si handle wafer can experience stronger heating and a larger thermomechanical response than the bonded diamond layer. This mismatch can generate interfacial stress and trigger local bond failure, especially when the buried oxide is only 300 nm thick and therefore cannot substantially accommodate the difference in expansion between the diamond and the Si carrier. Representative room-temperature thermal expansion coefficients are approximately $1.0$--$1.1$~ppm~K$^{-1}$ for diamond and $2.6$~ppm~K$^{-1}$ for single-crystal silicon, with the latter increasing further at elevated temperature \cite{nano14050460, egan2024expansivity-501, kelly2007composite-4b5}. This difference provides a credible route for plasma-induced interfacial stress during rapid heating and cooling.


To distinguish between these mechanisms, we prepared DOI samples with three controlled variations: SiO$_2$ thickness, SiO$_2$ deposition method, and carrier substrate material. Increasing the SiO$_2$ thickness reduces the likelihood that the buried oxide is fully removed by lateral attack from the edges, and is also advantageous for the later release of free-standing photonic structures. We compared PECVD and thermally grown SiO$_2$ because thick PECVD oxide can introduce higher surface roughness, which may complicate direct bonding, whereas thermally grown SiO$_2$ can provide micrometer-scale oxide thickness with high surface quality. Finally, fused quartz was introduced as an alternative carrier substrate to test the thermal-shock hypothesis. Its thermal expansion coefficient, approximately $0.5$--$0.55$~ppm~K$^{-1}$, is much closer to that of diamond than silicon~\cite{nano14050460, egan2024expansivity-501, kelly2007composite-4b5}, making quartz a useful platform for testing the thermal-shock hypothesis.


\subsection{Low-power etching using diamonds bonded to thick PECVD/ thermal oxide}


To reduce the risk of complete oxide loss during plasma processing, we increased the SiO$_2$ thickness from 300~nm to 2.5~$\mu$m using PECVD (Rq=4.44 nm).  Using this thick-oxide DOI substrate, we tested modified etching conditions with reduced ICP/RF power, and therefore lower thermal loading, as summarised in Table~\ref{etch_recipes}.

Thick-DOI gentle etch step 1 was a low-power version of the initial O$_2$-based diamond etch. After 15~min of processing, the diamond plate remained bonded, confirming that reducing the plasma load can mitigate the debonding observed under the initial high-power conditions. However, the reduced power also lowered the material-removal rate, making this condition insufficient by itself for practical DOI thinning. We therefore tested an oxygen-rich Ar/O$_2$ chemistry, denoted as Thick-DOI gentle etch step 2 in Table~\ref{etch_recipes}, where Ar was added to the O$_2$ plasma to increase the effective removal rate while avoiding the original high-power O$_2$ condition. The diamond remained bonded after 15~min of etching. The same Ar/O$_2$ recipe was then applied for an extended duration to a diamond plate bonded to 2.5~$\mu$m thermally grown SiO$_2$ on Si (Rq=0.2 nm), where the bonding interface was preserved even after 60~min of etching. These results show that Si-based DOI carriers can withstand extended diamond thinning when the plasma conditions are moderated.

Although the bonding was preserved, the surface quality after the extended Ar/O$_2$ etch was not suitable for photonic-grade DOI. Optical microscope and SEM inspections were therefore used to identify the origin of the roughening. As shown in Fig.~\ref{figure_sem_masking}a, the top-view optical microscope image reveals a high density of surface defects. After tilting the sample holder by 15$^\circ$, SEM inspection, shown in Fig.~\ref{figure_sem_masking}b,c, reveals a dense pillar-like morphology. These features are characteristic of micromasking-driven roughening, where local particles, residues or redeposited material partially shield the diamond during oxygen-based etching and lead  to the formation of nanopillars or pits~\cite{ruf_optically_2019,hicks_diamond_2019}. This observation is important for the process development: reducing the plasma power can preserve the direct-bonded DOI interface, but power reduction alone does not provide the surface quality required for photonic applications.



\begin{table*}[t]
\centering
\footnotesize
\renewcommand{\arraystretch}{1.12}
\setlength{\tabcolsep}{8pt}
\begin{threeparttable}
\caption{Optimized cyclic etching recipe developed in this work. One full cycle consists of 4~min Ar/Cl$_2$ followed by 4~min Ar/O$_2$.}
\label{tab:optimized_cyclic_recipe}
\begin{tabular*}{0.82\textwidth}{@{\extracolsep{\fill}}lcc@{}}
\toprule
\textbf{Parameter} & \textbf{Ar/Cl$_2$} & \textbf{Ar/O$_2$} \\
\midrule
ICP (W) & 100 & 200 \\
RF (W) & 250 & 300 \\
Gas flow (sccm) & Ar 15 / Cl$_2$ 7 & Ar 50 / O$_2$ 13 \\
Chamber pressure & 0.01 mbar & 0.01 mbar \\
Carrier temperature & 20$^\circ$C & 20$^\circ$C \\
Diamond etch rate (nm/min) & $\sim$10 & $\sim$50--55 \\
Step duration & 4 min & 4 min \\
\bottomrule
\end{tabular*}
\begin{tablenotes}[flushleft]
\footnotesize
\item The Ar/Cl$_2$ step is used primarily for cleaning/smoothing, while the Ar/O$_2$ step provides the main material removal.
\end{tablenotes}
\end{threeparttable}
\end{table*}




\subsection{High-power etching of diamond bonded to quartz carriers}

The moderated etching experiments on Si-based carriers showed that the bonding interface can be preserved once the plasma load is reduced, suggesting that complete under-etching of the buried oxide is unlikely to be the dominant cause of debonding. To directly test the role of thermomechanical mismatch, we prepared an additional DOI substrate by bonding diamond to a 300~nm PECVD SiO$_2$ layer on a fused-quartz carrier. Fused quartz was selected because its thermal expansion coefficient is much closer to that of diamond than silicon, making it a useful diagnostic substrate for testing the thermal-shock hypothesis~\cite{haismaDiversityFeasibilityDirect1994, kelly2007composite-4b5}.

After successful bonding, the quartz-based DOI substrate was etched using the original high-power reference conditions listed in Table~\ref{etch_recipes}, i.e., the same conditions that caused debonding on Si-based carriers. The sample remained bonded after both the 15~min O$_2$ etch and the subsequent 20~min Ar/Cl$_2$ step. This result strongly supports thermal expansion mismatch as the main origin of the debonding observed for diamond bonded to SiO$_2$/Si. At the same time, the quartz experiment also showed that bond survival alone is not sufficient for photonic-grade DOI thinning: severe micromasking was still observed after etching. Thus, the quartz carrier resolves the debonding problem in a diagnostic sense, but does not by itself provide a manufacturable route for DOI thinning on a conventional Si platform.

\subsection{Origin of particle-induced micromasking and process implications}

After resolving the debonding issue, micromasking-induced surface degradation became the remaining critical limitation for achieving photonic-grade DOI thinning. The responsible particles can originate from two routes. First, they may already be present on the diamond surface due to growth, polishing, handling, cleaning residues, or external contamination. Second, they may be generated during plasma processing when exposed regions of the surrounding carrier wafer or mask stack are etched and redeposited onto the diamond surface. During oxygen-based diamond etching, such local micromasks partially protect the underlying diamond and evolve into nanopillars or pits, as also reported for etched diamond membranes~\cite{ruf_optically_2019,hicks_diamond_2019}.

These observations led to two process changes. First, a fused-quartz top mask with a laser-cut opening was introduced to limit direct plasma exposure of the surrounding carrier surface and reduce redeposition-driven micromasking. Second, the Ar/Cl$_2$ cleaning step was incorporated cyclically rather than used only as a pre-etch. In this way, newly accumulated particles can be removed during long etches before they develop into large surface defects. Together with reduced handling time between cleaning, bonding, mounting, and etching, these changes define the process requirements for a manufacturable DOI-thinning route: the recipe must preserve the bonded interface while repeatedly refreshing the diamond surface during material removal.

\section{Final optimized cyclic recipe and manufactured photonic-grade DOI substrate}
\subsection{Etch parameters}

The recipe-development study above shows that a successful DOI-thinning process must satisfy three requirements simultaneously: the bonding interface must survive plasma exposure, carrier-induced redeposition must be minimized, and the diamond surface must be repeatedly cleaned during long etch runs. The final recipe developed in this work, therefore, combines moderated plasma conditions, carrier/mask engineering, and cyclic etching. In contrast to the initial high-power reference process, the main removal step is implemented using an Ar/O$_2$ plasma rather than pure O$_2$, while the Ar/Cl$_2$ plasma is retained as a dedicated cleaning and smoothing step. In addition, the carrier and mask configuration is designed to limit unnecessary exposure of the surrounding wafer surface, thereby reducing redeposition-driven micromasking.

The optimized cyclic recipe is summarized in Table~\ref{tab:optimized_cyclic_recipe}. One full cycle consists of a 4~min Ar/Cl$_2$ step followed by a 4~min Ar/O$_2$ step. The Ar/Cl$_2$ step primarily removes surface particulates and smooths the diamond surface, with an etch rate of approximately 10~nm/min under the selected conditions. The Ar/O$_2$ step provides the main diamond removal, with an etch rate of approximately 50--55~nm/min. Both steps are carried out at a chamber pressure of 0.01~mbar and a carrier temperature of 20$^\circ$C. Repeated execution of this 4~min + 4~min cycle enabled stable thinning of directly bonded diamond plates from the 10--20~$\mu$m range down to the photonic-grade regime. In the DOI demonstration reported here, a 10~$\mu$m bonded diamond plate was thinned to $\sim$300~nm over an area of $0.5 \times 0.5$~mm$^2$ while preserving the buried diamond--oxide interface. This optimized recipe completes the etch-development storyline. The earlier experiments showed that high-power reference recipes can trigger debonding on Si-based DOI, that lowering the plasma load preserves the bond but does not by itself suppress micromasking, and that quartz support confirms the role of thermal shock while also highlighting the need for improved particle control. The final cyclic Ar/Cl$_2$ + Ar/O$_2$ process addresses these issues together and is therefore used for the DOI substrates presented in the remainder of this work.


\begin{figure*}[!htbp]
	\centering 
	\includegraphics[width=1\textwidth]{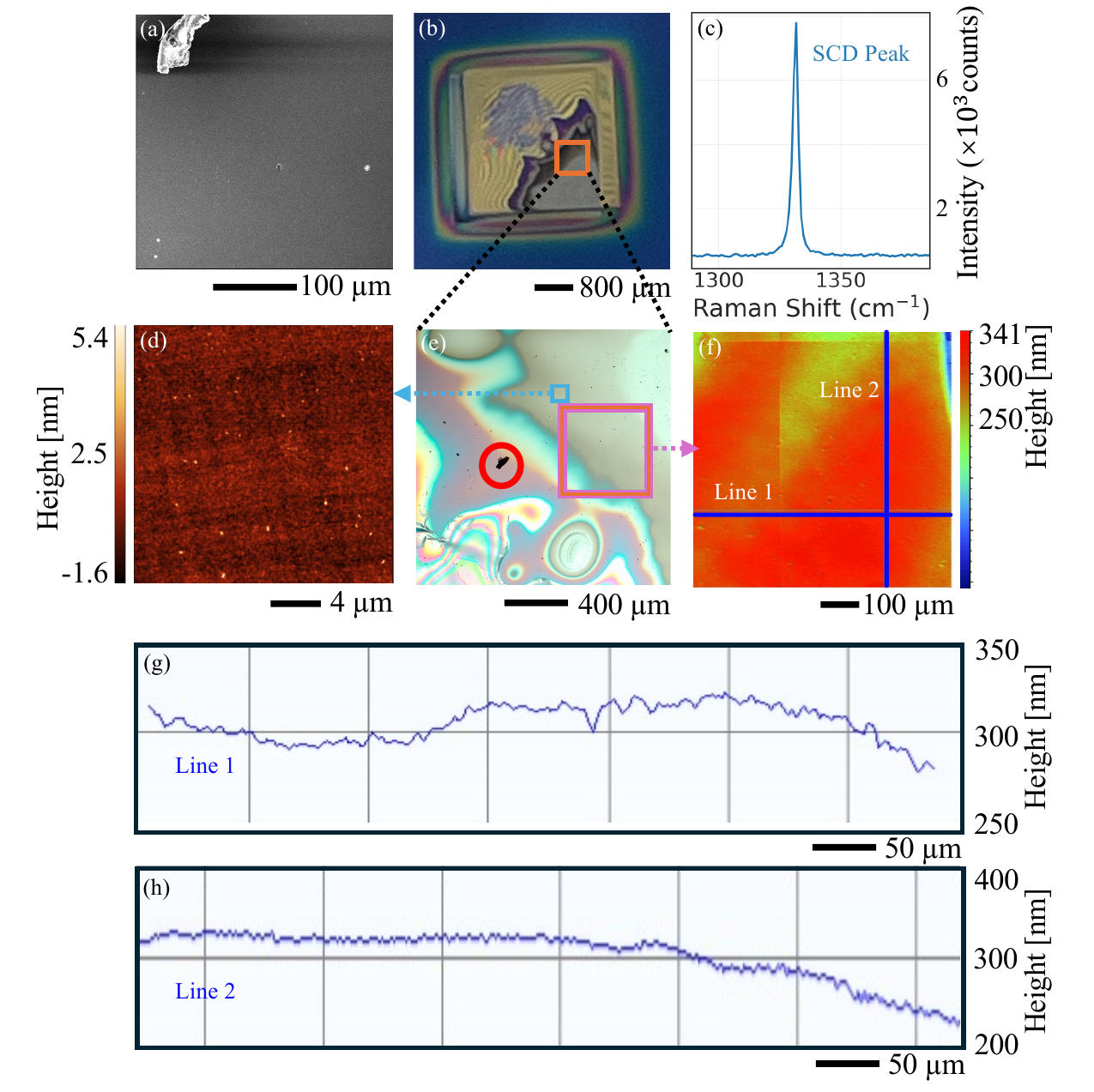}	
	\caption{Overall characterization of the fabricated DOI thin film. (a) SEM image at $\times$300 magnification. Sample surface does not have many visible contaminants, even though we didn't keep the entire etching/handling/inspection very clean. The large particle on the top right is used as a marker to localize the target area, for example, as the black dot in the red circle of subfigure (e) . (b) Optical microscopy image of the entire DOI substrate, where the orange square indicates the target $0.5\,\mathrm{mm} \times 0.5\,\mathrm{mm}$ region. (c) Raman spectrum from the target area which confirms the presence of SCD. (d) Representative AFM topography showing a roughness R$_q$ of $<$ 0.5 nm. (e) Zoomed-in optical microscopy image, and the color contrast corresponds to thickness variations, with the purple region is the target region. (f) WLI image of the target thin film. Thickness profiles were extracted along a horizontal line (line 1 in (g)) and a vertical line (line 2 in (h)), both indicating an average thickness of approximately 300 nm.} 
	\label{fig_all}
\end{figure*}




\begin{figure}[ht]
    \centering
    \vspace{1em}
    \includegraphics[width=0.5\textwidth]{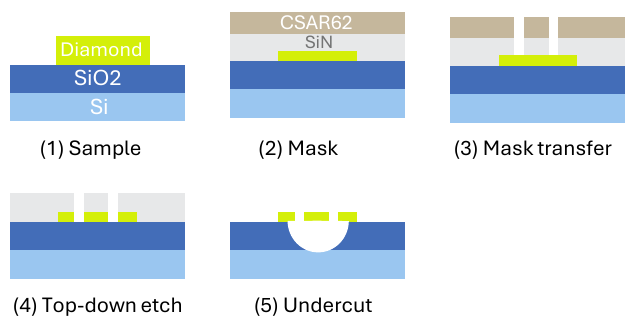}
    \caption{DOI based fabrication process. Sample is prepared from previous stage where sub-um thin diamond is directly bonded onto SiO$_2$/Si substrate via chemical bonds. Pattern is defined by E-beam lithography and and transferred via a SiN hard mask. HF undercut is used lastly for free-standing structures. More details can be found in the main text.}
    \label{process_flow_photonics}
    \vspace{1 em}
\end{figure}



\subsection{Photonic-grade DOI substrate} 

Using the optimized cyclic recipe, we achieved controlled thinning of a direct-bonded DOI substrate to the photonic-grade thickness range. Fig. ~\ref{fig_all} summarizes the characterization of the resulting thin film. The fabricated DOI contains a $0.5~\mathrm{mm} \times 0.5~\mathrm{mm}$ region with a target thickness of approximately 300~nm. We use the particle on top left of Fig. \ref{fig_all}a to localize the target area. As shown in Fig.~\ref{fig_all}b, the orange square marks the region reaching the target $\sim$300~nm thickness (verified by WLI measurements as shown in Fig. \ref{fig_all}f, 1D thickness profile data of two representative lines-horizontal (line 1) and vertical (line 2) are detailed in Fig. \ref{fig_all} g and h, respectively), while the surrounding bonded diamond remains thicker, up to approximately 1~$\mu$m. This spatial thickness profile arises from the mask-defined plasma exposure during etching and the initial thickness variation of the starting diamond plate. The target thin-film region is largely free of visible contaminants and micromasking-induced structures, confirmed by SEM images in Fig. \ref{fig_all}a. Surface characterization shows a roughness of $R_q < 0.5$~nm, shown in Fig. \ref{fig_all}d, indicating that the cyclic cleaning/etching strategy can preserve photonic-grade surface quality during deep thinning. We observe a $\sim 50 - 70$ nm thickness variation across the $0.5 \times 0.5$ mm$^2$ photonic-grade DOI area, which arises from the quartz mask-related etch non-uniformity and the initial thickness variation of the starting diamond plate. Importantly, the buried diamond--oxide interface remained intact after thinning, confirming that the optimized process can combine aggressive material removal with bond preservation and surface-quality control.

\subsection{Summary of etch process optimization} 

The recipe-development study leads to three main conclusions. First, the debonding observed with the original high-power reference recipes on directly bonded SiO$_2$/Si DOI is primarily associated with thermomechanical shock rather than complete under-etching of the buried oxide. This conclusion is supported by the survival of thick-oxide Si-based samples during moderated etching and by the successful high-power etching of diamond bonded to quartz. Second, preserving the bonding interface alone is not sufficient for photonic-grade DOI fabrication: long oxygen-rich etches can produce severe micromasking, nanopillars, and unacceptable roughening unless particle generation and redeposition are actively controlled. Third, a practical DOI-thinning process must combine moderated plasma conditions, carrier/mask engineering, and cyclic surface cleaning. 

Taken together, these findings define the optimized process used in this work: a quartz-assisted carrier/mask configuration combined with a 4~min Ar/Cl$_2$ + 4~min Ar/O$_2$ cyclic ICP-RIE recipe. The Ar/Cl$_2$ step repeatedly refreshes the surface by removing particulates and smoothing the diamond, while the Ar/O$_2$ step provides the etch rate required for efficient thinning. With this combined strategy, bonded SCD plates in the 10--20~$\mu$m range can be thinned to the photonic-grade regime. In the demonstrated process, a 10~$\mu$m bonded plate was reduced to a $\sim$300~nm DOI film over $0.5 \times 0.5$~mm$^2$ while preserving the buried interface and maintaining sub-nanometer surface roughness. This optimized cyclic recipe therefore provides the process foundation for the photonic-grade DOI substrates, chiplet fabrication, and thickness-validation results presented in the following sections.


\begin{figure*}[ht]
    \centering
    \includegraphics[width=0.7\textwidth]{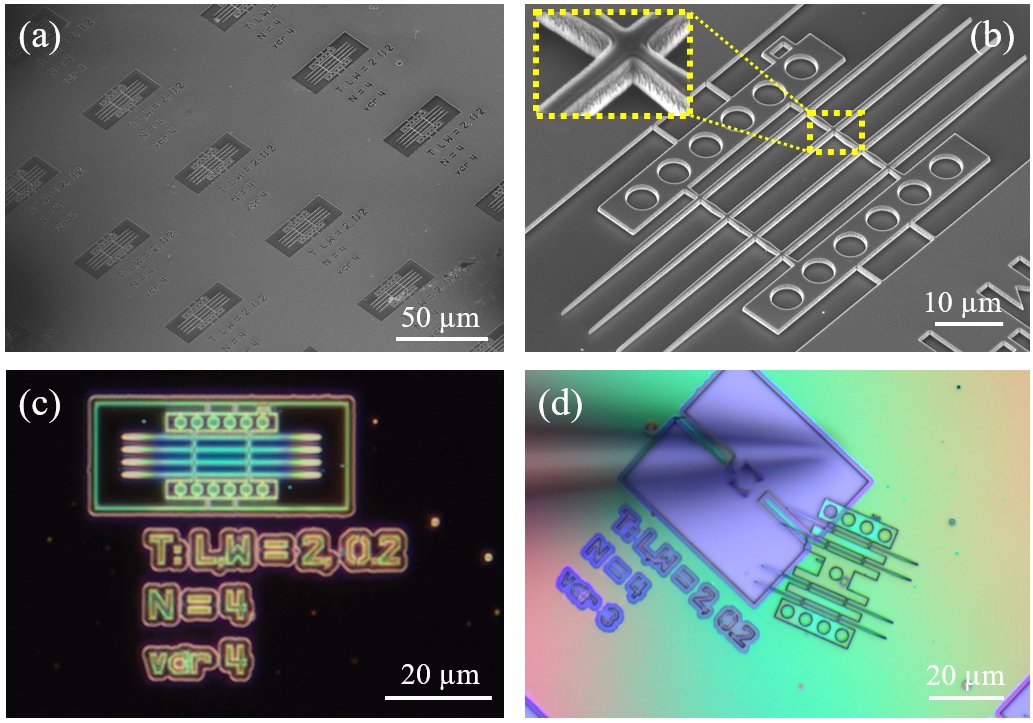}
    \caption{SEM and microscope inspection of fabricated chiplets. (a) Zoomed-out view of fabricated chiplets aray. (b) Zoomed-in image of a single chiplet. The tether structure and waveguides are fabricated successfully and connected by a supporting frame. The smooth top surface indicates low roughness after the whole frabrication process. The top-left inset image shows no pole structure underneath the crosses. (c) Dark-field optical microscope image of a chiplet together with (d) where a chiplet is picked up by a needle via van der Waals force indicate the successful undercut.}
    \label{figure_sem_chiplets}
\end{figure*}

\section{Diamond anisotropic etch recipe optimization for photonic chiplet fabrication from DOI substrate}



Device fabrication was carried out on the photonic-grade DOI substrate obtained in the previous section, consisting of a sub-$\mu$m-thick diamond layer bonded to a SiO$_2$/Si carrier wafer. The main fabrication steps are schematically illustrated in Fig. \ref{process_flow_photonics}. The substrate (step 1) was first cleaned by rinsing in acetone and isopropanol (IPA). In step (2), a SiN hard mask was then deposited by PECVD, followed by spin coating of AR-P-6200.09 (CSAR62) electron-beam resist. The chiplet pattern was defined in CSAR62 by electron-beam lithography and resist development, and subsequently transferred into the SiN hard mask using a CHF$_3$/O$_2$-based RIE (step 3) \cite{ruf_optically_2019, codreanu2025diamond-82a}. An oxygen-based anisotropic dry etch was then used to transfer the pattern into the diamond layer in step (4). Finally, in step (5) buffered oxide etchant (BOE, 7:1 v/v mixture of 40 wt$\%$ NH$_4$F and 49 wt$\%$ HF) at room temperature was used to remove the SiN mask and undercut the buried SiO$_2$. The sample was subsequently dried using critical point drying (CPD), thereby releasing the free-standing diamond structures from the DOI substrate.

\subsection{Anisotropic diamond etch on DOI substrate}
To transfer patterns from the SiN hard mask to the diamond substrate, a high-accuracy, top-down, anisotropic diamond etch process is required. For photonic device fabrication based on DOI substrate, two additional requirements must be satisfied. First, micro-masking during diamond etch must be addressed to ensure a smooth device surface. Second, the bonding interface must be preserved with high quality throughout the ICP-RIE process, which is also beneficial for the subsequent HF-based undercut step.

We first evaluated the cyclic diamond thinning-down recipe developed in Section \ref{thinning_down_recipe} which provides efficient anisotropic diamond etch while maintaining good surface quality and preserving the bonding interface. However, the recipe was optimized for anisotropic diamond etch using a quartz mask and was not directly compatible with pattern transfer using a SiN hard mask. The use of a SiN hard mask will cause micro-masking resulted from SiN material re-deposition during the Ar-atoms contained etch process. Therefore, a dedicated anisotropic diamond etch recipe had to be developed separately for pattern transfer from the SiN hard mask into the diamond substrate.

As an alternative starting point, we investigated a pure oxygen-plasma-based etching process \cite{MaxRuf_Network2021} with etch parameters listed in Table \ref{etch_recipes} ``Initial-DOI etch 1''. This etch process has shown highly anisotropic diamond etch, absence of micro-masking, and smooth sidewall of our target 200-300 nm device thickness range. However, the thermal-shock issue described in Section \ref{thinning_down_recipe} reappeared, even for etch durations of only several tens of seconds.

Based on the experience from these two processes, we developed a new cyclic diamond etch recipe tailored for top-down pattern transfer using a SiN hard mask. The recipe was designed to mitigate both thermal shock and micro-masking with a few key characteristics. First, a cyclic etch method is designed without heavy Ar ions and with reduced plasma power. Etch cycle loop consists of a 15 s O$_2$ etch step (50 sccm O$_2$, 10 $\mu$bar chamber pressure) using an RF power of 300 W and an ICP power of 200 W  on Oxford Instruments Plasmalab  System 100 RIE system. The etch step is followed by a 5 min cooling step. Second, the carrier-wafer table temperature was reduced from 20 $^\circ$C to 0 $^\circ$C to compensate for the heating induced by the high-power plasma. Third, the helium backside-cooling flow was increased from 5 sccm to 10 sccm to improve thermal contact between the DOI substrate and the carrier-wafer table.

\subsection{Photonic chiplets from DOI substrate}

Using the optimized etch recipes and process flow, we fabricated free-standing diamond photonic chiplets consisting of waveguides retained within a surrounding support frame, as shown in Fig.~\ref{figure_sem_chiplets}a,b. The frame provides mechanical robustness during release and handling, while the internal waveguide structures form the photonic element intended for heterogeneous integration. The suspension of the chiplet is confirmed by dark-field microscopy and by pick-and-place manipulation, as shown in Fig.~\ref{figure_sem_chiplets}c,d. In this test, a diamond chiplet is picked up with a probe through van der Waals adhesion, demonstrating compatibility with subsequent transfer onto a photonic integrated circuit receptor for hybrid photonic integration~\cite{wan2020largescale-f02, Leyla_PnP_prep}.

A higher-magnification SEM image of the waveguide cross section, shown in the inset of Fig.~\ref{figure_sem_chiplets}b, confirms that the waveguides are fully released and that no residual support pedestal remains underneath the structures. This is an important practical advantage of the DOI-based release process.

Here, we demonstrate a full-stack fabrication flow for a photonic-grade DOI substrate. Using this platform, we fabricate free-standing diamond photonic chiplets, while the same DOI approach also enables complex, non-suspended diamond photonic circuits directly supported by the oxide layer.



\section{Contrast and thickness evaluation of diamond layer on SiO\texorpdfstring{$_2$}{2} }

\begin{figure}[b!]
	\centering 
    \vspace{1em}
	\includegraphics[width=0.5\textwidth]{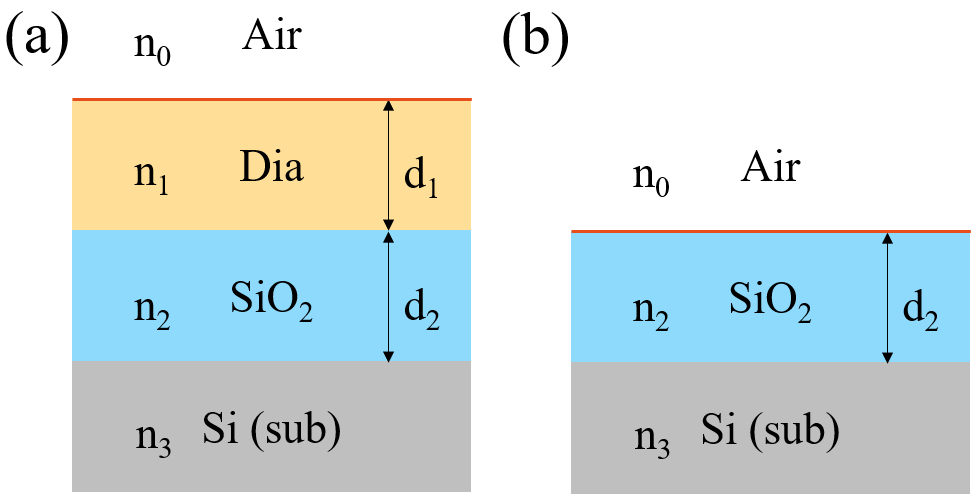}	
	\caption{(a) Trilayer scheme of diamond on SiO$_2$ / Si substrate and (b) bilayer scheme of the SiO$_2$ / Si substrate. } 
	\label{fig_mom0_fig6}%
\end{figure}

Under visible illumination, a DOI stack exhibits thickness-dependent colors due to thin-film interference in the diamond and SiO$_2$ layers. This effect can be used as a simple and rapid optical indicator of diamond thickness during thinning, similar to the optical-contrast methods widely used for graphene and other thin-layer systems on SiO$_2$/Si substrates~\cite{muller_visibility_2015, blake_making_2007, miranda_rendering_2017}. We model the DOI system as a four-layer stack composed of air, diamond, SiO$_2$, and Si, as schematically shown in Fig.~\ref{fig_mom0_fig6}a. The refractive indices are denoted as $n_0=1$ for air, $n_1=2.41$ for diamond~\cite{phillip_kramers-kronig_1964}, $n_2=1.46$ for SiO$_2$~\cite{herzinger_ellipsometric_1998, ghosh_dispersion-equation_1999}, and $n_3=4$ for Si~\cite{jellison_optical_1993}. The diamond thickness $d_1$ is varied from 0 to 1200~nm, while the SiO$_2$ thickness $d_2$ is varied from 0 to 600~nm. For this first-order contrast model, the refractive indices are taken at $\lambda=550$~nm and treated as constant across the visible range. This approximation captures the main interference-driven thickness dependence near the green channel of an optical microscope; full colour rendering can be obtained by extending the same transfer-matrix calculation over the visible spectrum with wavelength-dependent optical constants.

At normal incidence, the intensity reflectance of the DOI stack is calculated from the Fresnel coefficients and phase accumulation in the diamond and SiO$_2$ layers. Following the multilayer interference formalism used for thin-film visibility calculations~\cite{miranda_rendering_2017}, the reflectance of the air/diamond/SiO$_2$/Si stack is written as
\begin{equation}
    R_{\mathrm{Dia}} =
    \left|
    \frac{
    r_1 e^{i(\phi_1+\phi_2)}
    + r_2 e^{-i(\phi_1-\phi_2)}
    + r_3 e^{-i(\phi_1+\phi_2)}
    + r_1 r_2 r_3 e^{i(\phi_1-\phi_2)}
    }{
    e^{i(\phi_1+\phi_2)}
    + r_1 r_2 e^{-i(\phi_1-\phi_2)}
    + r_1 r_3 e^{-i(\phi_1+\phi_2)}
    + r_2 r_3 e^{i(\phi_1-\phi_2)}
    }
    \right|^2 ,
\end{equation}
where
\begin{equation}
    r_1=\frac{n_0-n_1}{n_0+n_1}, \qquad
    r_2=\frac{n_1-n_2}{n_1+n_2}, \qquad
    r_3=\frac{n_2-n_3}{n_2+n_3},
\end{equation}
and
\begin{equation}
    \phi_i = \frac{2\pi n_i d_i}{\lambda}
\end{equation}
is the phase accumulated in layer $i$ for wavelength $\lambda$.

For comparison, the reflectance of the bare SiO$_2$/Si substrate, shown schematically in Fig.~\ref{fig_mom0_fig6}b, is obtained by setting the diamond layer thickness to zero. The corresponding expression is
\begin{equation}
    R_{\mathrm{Sub}} =
    \left|
    \frac{
    r'_2 e^{i\phi_2} + r_3 e^{-i\phi_2}
    }{
    e^{i\phi_2} + r'_2 r_3 e^{-i\phi_2}
    }
    \right|^2 ,
\end{equation}
where
\begin{equation}
    r'_2=\frac{n_0-n_2}{n_0+n_2}.
\end{equation}

The optical contrast of a diamond film relative to a reference region can then be defined as the relative change in reflected intensity,
\begin{equation}
    C(\lambda)=
    \left|
    \frac{R_{\mathrm{Dia}}(\lambda)-R_{\mathrm{ref}}(\lambda)}
    {R_{\mathrm{Dia}}(\lambda)}
    \right| ,
\end{equation}
where $R_{\mathrm{ref}}$ can be either the reflectance of the bare SiO$_2$/Si substrate, $R_{\mathrm{Sub}}$, or the reflectance of a diamond region with a known reference thickness. This contrast definition provides a direct way to identify thickness ranges with high optical visibility and to select suitable SiO$_2$ thicknesses for microscope-based thickness evaluation. In the following analysis, we use this interference model to calculate the expected color and contrast evolution of DOI stacks and then correlate the calculated trends with optical micrographs and white-light interferometry measurements.


\subsection{Contrast of diamond for single wavelength}

\begin{figure*}[ht]
	\centering 
	\includegraphics[width=1\textwidth]{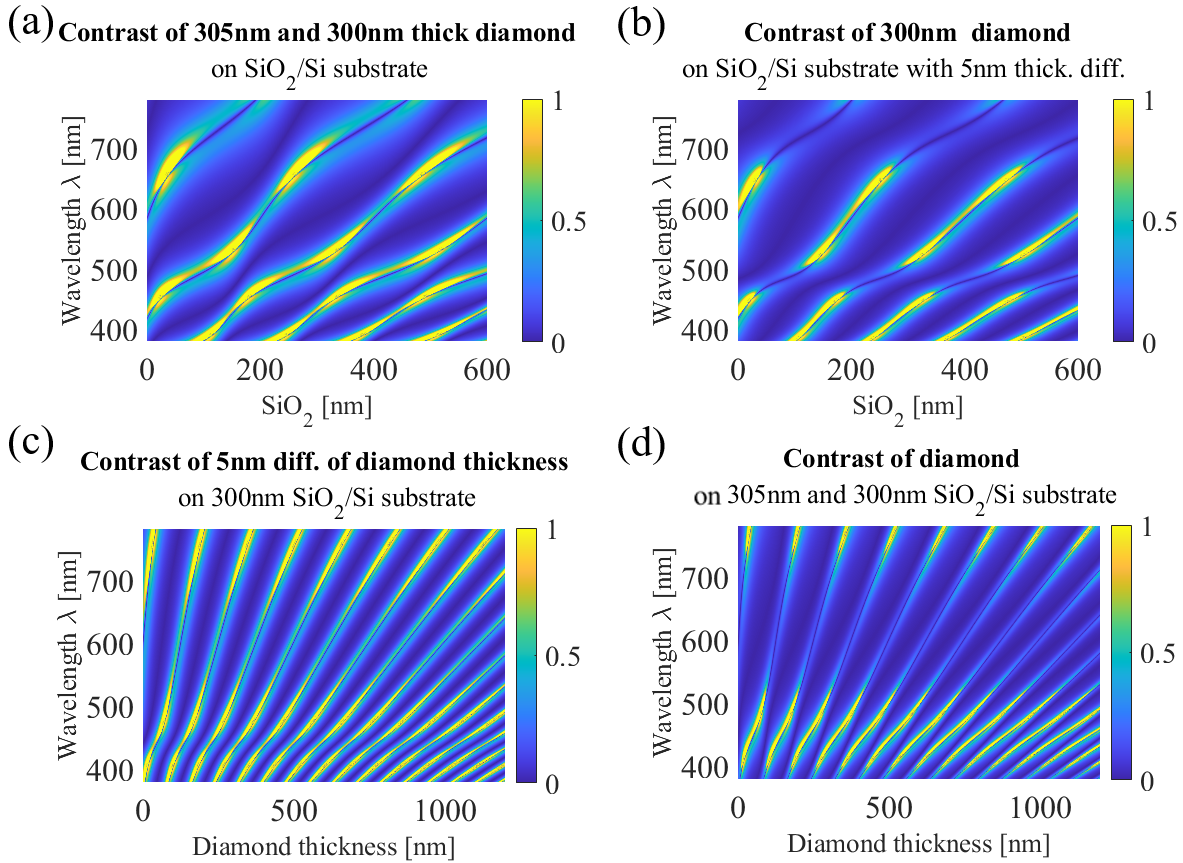}	
	\caption{Contrast between (a) diamond layers with 305nm and 300nm thickness on  SiO$_2$/Si with thickness of  SiO$_2$ varying from 0 to 600nm , (b) 300nm thick diamond layers on varying SiO$_2$/Si substrates differing of 5nm SiO$_2$ thickness, (c) diamond layers differing of 5nm thickness on a 300nm SiO$_2$/Si substrate for diamond thicknesses varying from 0 to 1µm (d) diamond layers of the same thickness on 305nm and 300nm   SiO$_2$/Si substrates for diamond thicknesses varying from 0 to 1µm.} 
	\label{fig_mom0_fig7}%
\end{figure*}

\begin{figure*}
	\centering 
	\includegraphics[width=1\textwidth]{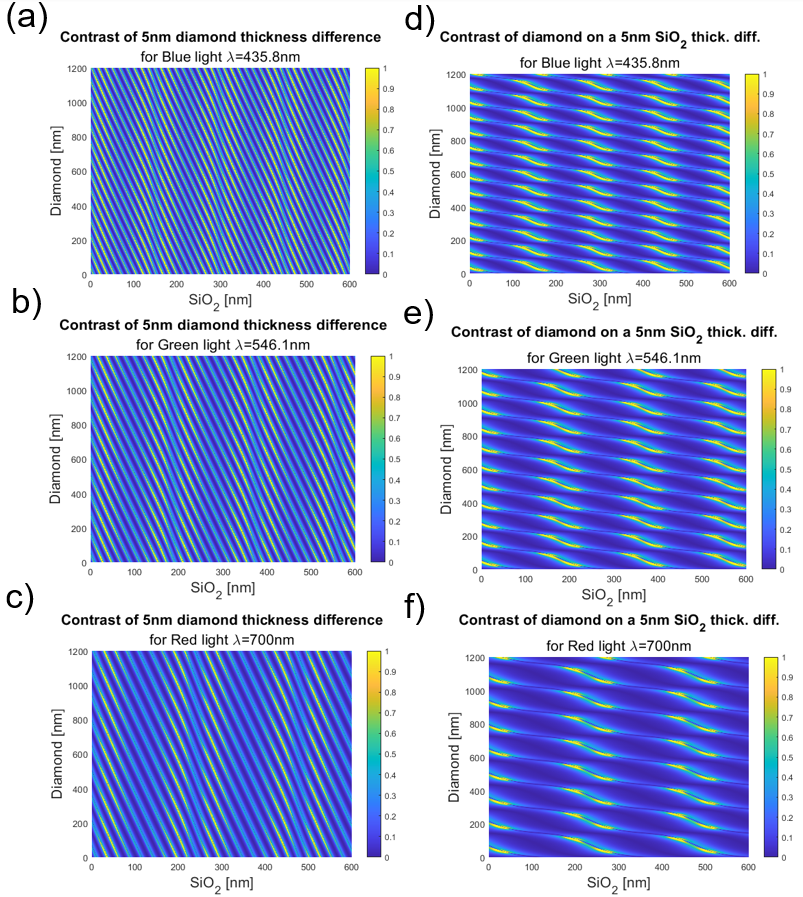}	
	\caption{Contrast of diamond layers with 5nm thickness difference on SiO$_2$ substrate for monochromatic a) blue, b) green, and c) red light; contrast of diamond on SiO$_2$ substrate with 5nm thickness difference for monochromatic d) blue, e) green, and f) red light. } 
	\label{fig_mom0_fig08}%
\end{figure*}

As shown by the reflectance expressions above, the optical contrast of a DOI stack depends on the thicknesses and refractive indices of the constituent layers, as well as on the illumination wavelength. In Fig.~\ref{fig_mom0_fig7}a, the SiO$_2$ thickness is varied from 0 to 600~nm while the contrast is calculated between diamond layers of 300 and 305~nm thickness. In this case, the contrast shows a pronounced modulation with broader spacing between maxima. A comparable trend is observed in Fig.~\ref{fig_mom0_fig7}b, where a 300~nm diamond layer is placed on SiO$_2$ layers differing by 5~nm in thickness. 

Fig.~\ref{fig_mom0_fig7}c shows the calculated contrast between a diamond layer of thickness $d$ and a layer 5~nm thicker, both on a 300~nm SiO$_2$/Si substrate, for wavelengths in the visible range (380--780~nm) and diamond thicknesses from 0 to 1200~nm. For a fixed wavelength, the contrast exhibits a series of maxima. These maxima are approximately periodic in diamond thickness, and their spacing increases with wavelength, as expected from the phase accumulation in the diamond film. A similar behavior is obtained when the oxide thickness is varied. Fig.~\ref{fig_mom0_fig7}d shows the contrast between two otherwise identical DOI stacks in which the SiO$_2$ thicknesses are 300 and 305~nm, respectively. The resulting pattern is qualitatively similar to that in Fig.~\ref{fig_mom0_fig7}c, indicating that small changes in either the diamond or oxide thickness can produce measurable changes in optical contrast. 

The maxima in Fig.~\ref{fig_mom0_fig7} identify thickness and wavelength combinations for which small changes in diamond or SiO$_2$ thickness produce the largest optical response. These conditions therefore provide useful guidance for selecting oxide thicknesses and illumination wavelengths for thickness evaluation. Physically, the maxima arise from coupled thin-film interference in the diamond and SiO$_2$ layers. The SiO$_2$ layer is bounded by higher-index media in the DOI stack, whereas the diamond layer is bounded by lower-index media at the air/diamond and diamond/SiO$_2$ interfaces. The resulting phase conditions act as coupled Fabry--Pérot-like resonances that enhance the sensitivity of reflectance to small thickness variations. Fig.~\ref{fig_mom0_fig08} shows the contrast as a function of diamond and SiO$_2$ thickness under monochromatic blue ($\lambda=435.8$~nm), green ($\lambda=546.1$~nm), and red ($\lambda=700$~nm) illumination. In Fig.~\ref{fig_mom0_fig08}a--c, the contrast is calculated between diamond layers differing by 5~nm. Two families of contrast maxima are visible, corresponding to the interference conditions associated with the diamond and SiO$_2$ layers. The spacing between these features increases with wavelength. A similar pattern of overlapping maxima is observed in Fig.~\ref{fig_mom0_fig08}d--f, where the compared systems differ by 5~nm in SiO$_2$ thickness while the diamond thickness is kept fixed. These calculations show that both diamond and oxide thickness variations can be detected optically, but that sensitivity depends strongly on wavelength and local position in the DOI thickness map.


\subsection{Contrast of diamond in the visible spectrum}

The single-wavelength analysis is directly relevant for filtered or monochromatic illumination. However, optical microscope images acquired under standard white-light conditions contain contributions from the full visible spectrum. To describe the perceived color of the DOI stack, we used the colorimetric CIE 1931
XYZ space~\cite{schanda_colorimetry_2007} and calculated the XYZ tristimulus values as
\begin{equation}
\label{XYZ_Values}
\left(
\begin{array}{c}
X \\
Y \\
Z
\end{array}
\right)
=
P \int_{380~\mathrm{nm}}^{780~\mathrm{nm}}
S(\lambda) R(\lambda)
\left(
\begin{array}{c}
\overline{x}(\lambda) \\
\overline{y}(\lambda) \\
\overline{z}(\lambda)
\end{array}
\right)
d\lambda ,
\end{equation}
where $\overline{x}(\lambda)$, $\overline{y}(\lambda)$, and $\overline{z}(\lambda)$ are the CIE 1931 color-matching functions tabulated in Ref.~\cite{schanda_colorimetry_2007}, $R(\lambda)$ is the calculated reflectance of the DOI stack, and $S(\lambda)$ is the spectral power distribution of the illumination source. For the microscope illumination used here, the source is approximated by CIE Illuminant A, corresponding to a tungsten-filament source with a colour temperature of approximately 2856~K. The prefactor $P$ accounts for the overall intensity normalization.

\begin{figure}
	\centering 
	\includegraphics[width=0.5\textwidth]{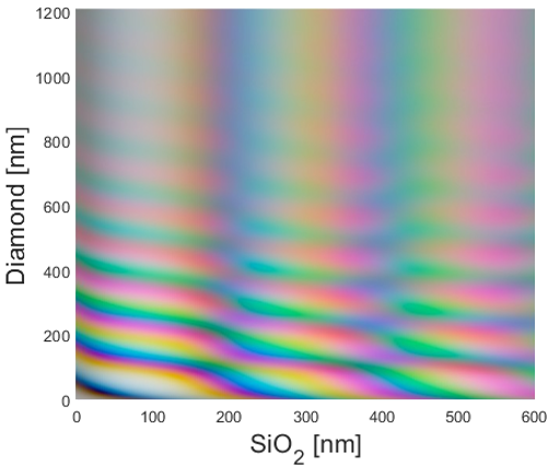}	
	\caption{Colour map of diamond on a SiO$_2$/Si substrate for assuming a brightness of 100\%.} 
	\label{fig_mom0_fig09}%
    \vspace{1em}
\end{figure}

The calculated XYZ values can be converted to RGB using standard color-space transformations, followed by display-specific corrections such as gamma correction for sRGB or AdobeRGB. This procedure produces a simulated color map of the DOI stack from its calculated reflectance spectrum. Fig.~\ref{fig_mom0_fig09} shows such a color map for diamond thicknesses from 0 to 1200~nm and SiO$_2$ thicknesses from 0 to 600~nm, displayed in the sRGB color space for brightness of 100\% or $P=1$. Changing $P$ changes the calculated luminance and therefore the displayed brightness; in the absence of clipping or display nonlinearities, the chromaticity is preserved. The color map shows that the DOI stack provides strong thickness-dependent color contrast, especially for thinner diamond and SiO$_2$ layers. Although the colour response is not globally one-to-one because thin-film interference is periodic, sufficiently distinct colour regions can be used for rapid thickness estimation when combined with an appropriate calibration.

For a quantitative comparison between two colours, it is useful to transform the XYZ values into the CIE 1976 $L^\ast a^\ast b^\ast$ colour space, which is more perceptually uniform than RGB or XYZ~\cite{schanda_colorimetry_2007, gao_total_2008}. The CIELAB coordinates are obtained from

\begin{equation} 
\label{eq:cielab_transform}
\begin{aligned}
L^\ast &= 116 f\left(\frac{Y}{Y_n}\right)-16, \\
a^\ast &= 500 \left[
f\left(\frac{X}{X_n}\right) -
f\left(\frac{Y}{Y_n}\right)
\right], \\
b^\ast &= 200 \left[
f\left(\frac{Y}{Y_n}\right) -
f\left(\frac{Z}{Z_n}\right)
\right],
\end{aligned}
\vspace{1 em}
\end{equation} 

where $X_n$, $Y_n$, and $Z_n$ are the tristimulus values of the reference white point. For Illuminant A and the CIE 1931 2$^\circ$ standard observer, these values are $X_n=109.850$, $Y_n=100$, and $Z_n=35.585$. The function $f(t)$ is defined as
\begin{equation}
f(t)=
\left\{
\begin{array}{ll}
t^{1/3}, & \mathrm{if}~t>\left(\frac{6}{29}\right)^3, \\[4pt]
\frac{1}{3}\left(\frac{29}{6}\right)^2 t+\frac{4}{29}, & \mathrm{otherwise}.
\end{array}
\right.
\end{equation}

In this space, $L^\ast$ represents perceptual lightness, from 0 for black to 100 for white, while $a^\ast$ represents the green--red axis and $b^\ast$ represents the blue--yellow axis. Unlike RGB-based color spaces, CIELAB is device independent once the illuminant and reference white point are specified. It is therefore better suited for comparing colors generated by different DOI thicknesses.

\begin{figure}
	\centering 
	\includegraphics[width=0.5\textwidth]{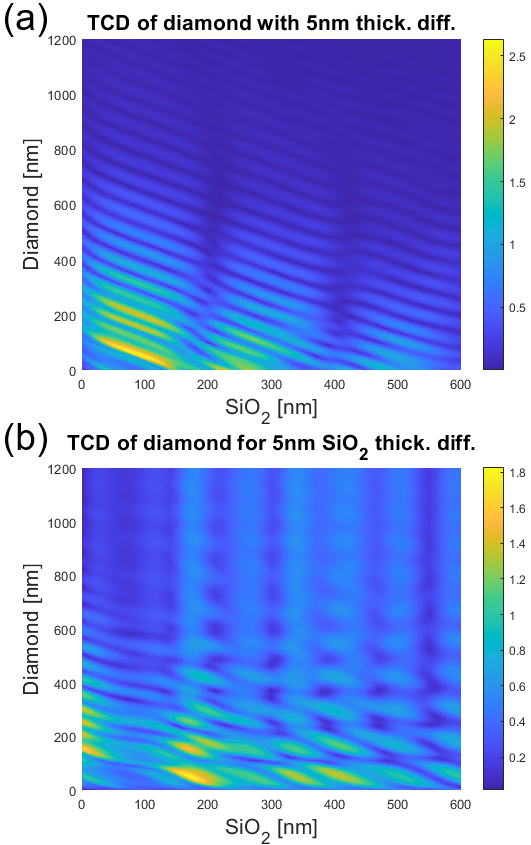}	
	\caption{Total Color Difference (TCD) for (a) diamond layers with 5nm thickness difference, (b) diamond layers with same thickness laying on SiO$_2$ substrate with 5nm layer thickness.} 
	\label{fig_mom0_fig10}%
    \vspace{1 em}
\end{figure}

The total color difference (TCD), expressed here using the CIE76 color-difference metric
\(\Delta E_{ab}^{\ast}\), quantifies the perceptual distance between two reflected colors in the CIE \(L^{\ast}a^{\ast}b^{\ast}\) color space. The TCD is then calculated using the CIE76 Euclidean distance,
\begin{equation}
TCD_{\mathrm{CIE76}} =
\sqrt{
\left(\Delta L^\ast\right)^2+
\left(\Delta a^\ast\right)^2+
\left(\Delta b^\ast\right)^2
}.
\end{equation}
This metric provides a simple quantitative measure of perceptual color separation between two DOI stacks. Figure~\ref{fig_mom0_fig10}a shows the total color difference between two systems whose diamond thicknesses differ by 5~nm, while Fig.~10b shows the corresponding result for a 5~nm difference in SiO$_2$ thickness. The TCD map in Fig.~\ref{fig_mom0_fig10}a follows the main color variations visible in Fig.~\ref{fig_mom0_fig09}, confirming that CIELAB distance is a useful metric for identifying thickness regions with high optical sensitivity. Figure~\ref{fig_mom0_fig10}b shows additional modulation features associated with the oxide-layer interference. A full quantitative interpretation of these secondary features is beyond the scope of this work; here, the TCD analysis is used primarily as a practical guide for selecting DOI thickness ranges that provide strong color sensitivity for microscope-based thickness evaluation.


\subsection{Algorithm for the thickness evaluation of diamond layer on SiO\texorpdfstring{$_2$}{2} }
Based on the above considerations, we propose an easy algorithm to extrapolate the thickness of the diamond layer in a DOI substrate using colorimetric analysis.
The algorithm consists of two parts: the first part is used for calibrating brightness on a sample with known thicknesses, and the second part is used to extrapolate the thickness of the diamond on the DOI under test using the previously-calibrated brightness.

\begin{figure}[t!]
	\centering 
	\includegraphics[width=0.5\textwidth]{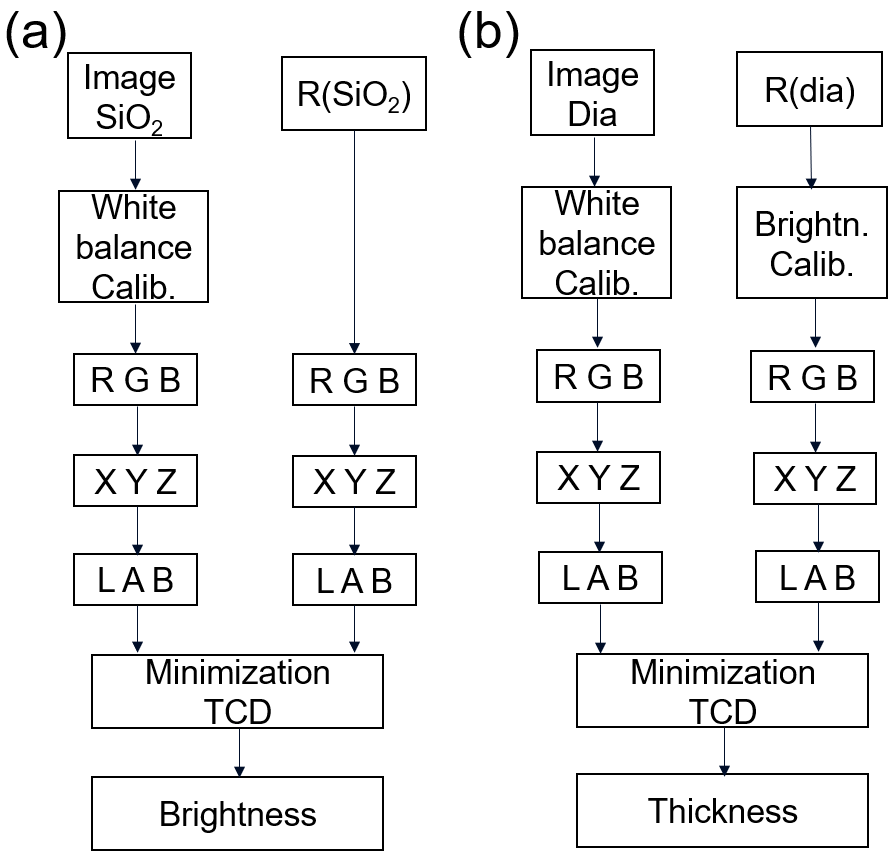}	
	\caption{Schematics of the colorimetric algorithm used for the extrapolation of the diamond thickness from an image: (a) first part of the algorithm in which the brightness of the image of a reference sample with known thickness is extrapolated by minimizing the TCD between the colors of the image  and the calculated ones for the same system; (b)  second part of the algorithm where the thicknesses of diamond in selected points of an image are extrapolated by minimizing the TCD of their color image and the calculated ones with the brightness calculated in (a).} 
    \vspace{1 em}
	\label{fig_mom0_fig11}%
\end{figure}

With reference to Fig.~\ref{fig_mom0_fig11}, the first part of the algorithm begins with an image taken of a substrate whose thicknesses are at least approximately known. Ideally, this should be the same substrate on which the diamond layer is placed (bonded). This image is used to estimate the brightness of the system. The image is taken with a known illuminant, for example, the illuminant A for an incandescent lamp, as in standard microscopes, and without any optical filters. After white balance adjustment, the RGB values for this image are converted into XYZ coordinates using the relationships in the literature \cite{noauthor_httpwwwbrucelindbloomcom_nodate, schanda_colorimetry_2007}. These transformations also account for the gamma correction associated with the monitors' sRGB or Adobe RGB color spaces. The XYZ coordinates are then converted into CIE LAB (\(L^\ast a^\ast b^\ast\)) coordinates using Eq.~\eqref{eq:cielab_transform}. Since the thickness of the reference system is known, the LAB coordinates can also be calculated theoretically. This is done by sweeping the brightness parameter (P) in Eq.~\eqref{XYZ_Values}, and, if needed, by varying the thickness around its known value. The source brightness, and at the same time the reference thickness, are then obtained from the condition that minimizes the total color difference (TCD) between the measured and calculated colors.

\begin{figure}[t!]
	\centering 
    \vspace{1 em}
	\includegraphics[width=0.4\textwidth]{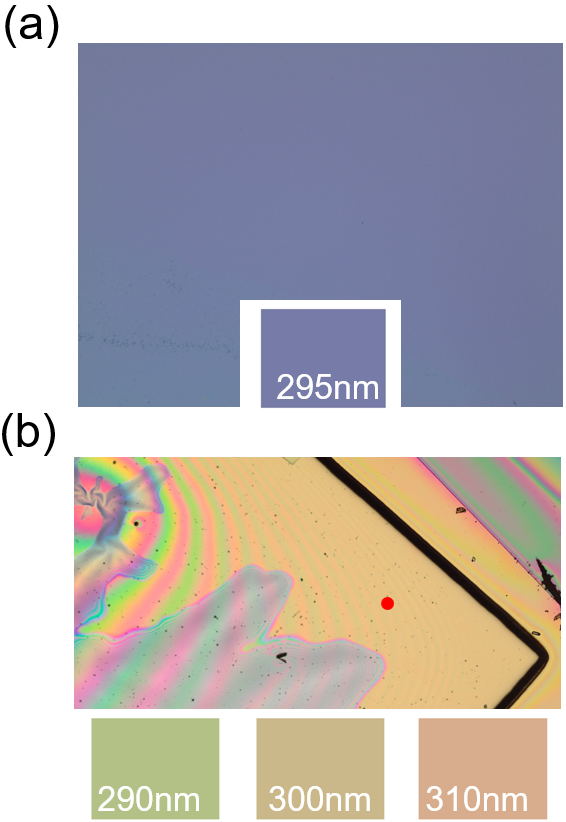}	
	\caption{Example of application of the method: (a) image of a sample with a known thickness (295nm thick SiO$_2$ without diamond, 5X objective, halogen lamp), and, in the frame,  calculated image minimizing the TCD with corresponding to the same thickness and a brightness of 84\%, (b) image of a sample of which we want to extrapolate the thickness on diamond (located at the red dot) taken with the same condition and in the, frames, simulated colors for SiO$_2$ / Dia=295nm / 290nm, SiO$_2$ / Dia=295nm / 300nm, the closes in color among the threee, and, SiO$_2$ / Dia=295nm / 310nm.} 
	\label{fig_mom0_fig12}%
    \vspace{1 em}
\end{figure}

In the second part of the algorithm, an image of the DOI sample under test is acquired under the same optical conditions used in the first part. The white balance is adjusted in the same way, and the RGB values are extracted from the points of interest. These values are converted into XYZ coordinates, calibrated using the brightness parameter (P) obtained in the first part of the algorithm, and then transformed into CIE LAB coordinates (\(L^\ast a^\ast b^\ast\)). Next, the system's reflectivities are calculated over a range of diamond thicknesses. The calculated reflectivity spectra are converted into XYZ coordinates and then transformed into LAB coordinates. The thickness of the diamond, which minimizes the TCD between the calculated and experimentally extrapolated coordinates, is assumed to be the effective thickness of the diamond layer.

Fig.~\ref{fig_mom0_fig12} shows a concrete example of thickness extrapolation for a DOI sample with nominal  diamond/SiO$_2$ thicknesses of 300 nm/300 nm
The reference sample, shown in Fig.~\ref{fig_mom0_fig12}a, is a SiO$_2$/Si substrate without diamond and with a known SiO$_2$ thickness of 300 nm. The DOI substrate with diamond is shown in Fig.~\ref{fig_mom0_fig12}b where the red dot indicates the point of interest. The images were taken with a camera (Olympus SC30), attached to an optical microscope (Olympus BX51) with a 5X objective (Olympus MPLFLN 5X) of numerical aperture 0.15 and 2.65 mm of field of view. The exposure time was kept the same for both images. Because of the large field of view and the low numerical aperture, the incident light can be approximated as straight, with negligible angular spread. 

In the frame of Fig.~\ref{fig_mom0_fig12}, we show the color obtained by minimization of the TCD. In this case, we varied the brightness and, in a small range, the thickness. In this case, the minimal TCD is obtained for a brightness of 84\% and a SiO$_2$ thickness of 295nm, quite close to the nominal thickness of 300nm. Performing the second part of the algorithm with this brightness and SiO$_2$ thickness, we can similarly find that the thickness of the diamond at the point of interest is 304 nm. By comparing the thickness extrapolated from our analysis with the corresponding thickness measured by the WLI, we can assess that our method estimates the correct thickness within a range of $\pm3$nm.  This is remarkable considering the simplicity of the method, which can provide a good initial estimate of the diamond thickness ~\cite{smith_color_1978}.%
To give a visual comparison of the color difference between similar diamond thicknesses, we added three frames at the bottom of Fig.~\ref{fig_mom0_fig12}b. These frames show, from left to right, the calculated colors for diamonds of thicknesses 290 nm, 300 nm, and 310 nm on the same substrate. The color that most closely resembles the diamond color in the image corresponds to 300 nm, which is also the closest value to the extrapolated diamond thickness. The process can also be automated to produce a thickness map over a full region of interest \cite{nolen_high-throughput_2011}.

It is worth noting that this is a simple model, and a perfect match between the image and the simulation cannot be fully expected. Several factors are not completely controllable. First, the calibration of the light source and the white balance may not be ideally described by the chosen illuminant \cite{muller_visibility_2015, ouyang_optical_2013}. The refractive indices of the investigated materials may also differ from the literature values. In addition, the color transformations applied by the digital microscope hardware and software may not be perfectly described by the chosen color space.
Second, the finite numerical aperture of the objective lens means that the incident light is not perfectly perpendicular to the sample, as assumed in the simulations \cite{roddaro_optical_2007}. These incident angles usually follow a Gaussian distribution \cite{jung_colors_2012, ouyang_optical_2013, gao_total_2008}, with a maximum incident angle $\theta\ \textsubscript {max}\ = \ \arcsin(NA)$ in air. The corresponding optical paths produce different phases and, therefore, modify the reflectivity of the material. However, this effect is expected to be low or reduced for thinner films \cite{gao_total_2008}, and for the low numerical aperture used in this work.

\section{Summary and conclusion}

In this work, we developed a practical route for producing photonic-grade DOI substrates from directly bonded single-crystal diamond on SiO$_2$/Si. The main challenge was not only to remove several micrometers of diamond, but to do so without breaking the bonded interface or damaging the surface needed for low-loss nanophotonics. By comparing different plasma conditions and carrier configurations, we found that high-power etching can trigger debonding primarily through thermomechanical stress, whereas long oxygen-rich etches can lead to severe micromasking and nanopillar formation. These limitations were overcome by combining a quartz-assisted carrier/mask configuration with a cyclic ICP-RIE process consisting of Ar/Cl$_2$ cleaning/smoothing and Ar/O$_2$ diamond removal. Using this process, a 10~$\mu$m directly bonded diamond plate was thinned to an approximately 300~nm DOI film over a $0.5 \times 0.5$~mm$^2$ region, while maintaining an intact buried interface and sub-nanometer surface roughness.

The fabricated DOI film was further used to demonstrate free-standing frame-supported diamond photonic chiplets, showing that the substrate is compatible with standard lithography, dry etching, oxide release, and pick-and-place handling. In parallel, we developed a colorimetric thickness-evaluation method based on thin-film interference in the diamond/SiO$_2$/Si stack, enabling rapid thickness estimation from standard optical microscope images with about 5~nm resolution and good agreement with WLI measurements. Together, the optimized thinning process, chiplet demonstration, and simple optical metrology form a compact manufacturing platform for scalable diamond nanophotonics. This approach provides a direct path from bonded diamond plates to device-ready DOI substrates and can support scalable integration of diamond quantum photonic circuits, heterogeneous chiplets, and broader diamond-based photonic and electronic systems.

\section*{Acknowledgements}

We gratefully acknowledge support from the joint research program “Modular quantum computers” by Fujitsu Limited and Delft University of Technology, co-funded by the Netherlands Enterprise Agency under project number PPS2007.

\appendix


\bibliographystyle{elsarticle-num}
\bibliography{References.bib,Colorimetry.bib}

\end{document}